\documentclass[11pt,psfig]{article}
\usepackage{graphicx}
\usepackage{amsmath,amssymb,mathtools}
\usepackage{bbold}
\usepackage{color}
\usepackage{cite}
\usepackage{tikz}
\usetikzlibrary{shapes}
\usepackage{longtable}
\usepackage{pgfplots}
\usepackage{subeqnarray}
\usepackage{dsfont}
\usepackage{changepage}

\tikzset{cross/.style={cross out, draw=black, minimum size=2*(#1-\pgflinewidth), inner sep=0pt, outer sep=0pt},
cross/.default={1pt}}

\title{Extending the range of validity of Fourier's law into the kinetic transport regime via asymptotic solution of the phonon Boltzmann transport equation}
\author{Jean-Philippe M. P\'{e}raud and Nicolas G. Hadjiconstantinou, \\Department of Mechanical Engineering, Massachusetts Institute of Technology\\ Cambridge, MA  02139, USA}

\begin{document}

\maketitle    

\begin{abstract}
We derive the continuum equations and boundary conditions governing phonon-mediated heat transfer in the limit of small but finite mean free path from asymptotic solution of the linearized Boltzmann equation in the relaxation time approximation. Our approach uses the ratio of the mean free path to the characteristic system lengthscale, also known as the Knudsen number, as the expansion parameter to study the effects of boundaries on the breakdown of the Fourier descrition. We show that, in the bulk, the traditional heat conduction equation using Fourier's law as a constitutive relation is valid at least up to second order in the Knudsen number for steady problems and first order for time-dependent problems. However, this description does not hold within distances on the order of a few mean free paths from the boundary; this breakdown is a result of kinetic effects that are always present in the boundary vicinity and require solution of a Boltzmann boundary-layer problem to be determined. Matching the inner, boundary layer, solution to the outer, bulk, solution yields boundary conditions for the Fourier description as well as additive corrections in the form of {\it universal} kinetic boundary layers; both are found to be proportional to the bulk-solution gradients at the boundary and parametrized by the  material model and the phonon-boundary interaction model (Boltzmann boundary condition). Our derivation shows that the traditional no-jump boundary condition for prescribed temperature boundaries and no-flux boundary condition for diffusely reflecting boundaries are appropriate only to zeroth order in the Knudsen number; at higher order, boundary conditions are of the jump type. We illustrate the utility of the asymptotic solution procedure by demonstrating that it can be used to predict the Kapitza resistance (and temperature jump) associated with an interface between two materials. All results are validated via comparisons with low-variance deviational Monte Carlo simulations.

\end{abstract}

%
\section{Introduction}

Microscale and nanoscale solid state heat transfer as mediated by phonon transport has received considerable attention in connection with a number of diverse practical applications, such as heat management in microelectronic devices, passive cooling and thermoelectric energy conversion \cite{Cahill2014}, but also due to the number of scientific challenges it poses. Particularly notable is the wide range of scales present in these problems, typically starting from the atomistic (including quantum) and extending to the macroscopic (device).  Kinetic-theory approaches based on the Boltzmann transport equation (BTE) \cite{Ziman1960}, especially if informed by ab-initio information on the material properties \cite{Landon2014c,Broido2007,Li2012}, can be quite effective in bridging this range of scales. One limitation of such approaches appears in the small mean free path limit,  $\langle \text{Kn} \rangle \ll 1$, where kinetic descriptions become stiff. Here, $\langle \text{Kn} \rangle$ denotes the Knudsen number defined as the ratio of the mean free path to the characteristic system lengthscale; a more precise definition will be given in section \ref{background}. 

As is well known, in the limit $\langle \text{Kn} \rangle\rightarrow 0$,  the stiff Boltzmann description need not be used because it can be replaced by the heat conduction equation; derivation of the bulk thermal conductivity from the Boltzmann equation in the relaxation approximation via a Chapman-Enskog type of expansion \cite{vincenti,Peraud2014} is well established, thus providing a ``pathway'' for recording the effect of molecular structure on the constitutive behavior in that limit. However, the Chapman-Enskog expansion is only applicable in the bulk and provides no information on the boundary conditions that need to supplement the heat conduction description in order to obtain solutions that are consistent with the (more fundamental) Boltzmann solution. Moreover, a rather large gap exists between lengthscales that truly satisfy $\langle \text{Kn} \rangle \rightarrow 0$ and the regime where Boltzmann equation solution is no longer problematic ($\langle \text{Kn} \rangle \gtrsim 0.1$).   

In this paper, we use an asymptotic expansion procedure using $\langle \text{Kn} \rangle$ as a small parameter to derive, from the BTE, the ``continuum'' equations governing phonon-mediated heat transfer in the small mean free path limit. This procedure recovers the classic heat conduction equation (including Fourier's law as a constitutive relation)  as the equation governing the temperature field that is consistent with solution of the Boltzmann equation to order $\langle \text{Kn} \rangle^0$, as expected. However, in contrast to Chapman-Enskog-type procedures, this procedure, also {\it derives}  the boundary conditions that the heat equation is to be solved subject to. Specifically, for fixed temperature boundaries, the Fourier boundary conditions are found to be of the Dirichlet type at the boundary temperature; for diffusely specular walls, the Fourier boundary conditions are shown to be the Neumann no-flux boundary condition. Although these results have been {\it empirically} established centuries ago, this is the first time they are shown to arise, rigorously, from a solution of the Boltzmann equation. 

More importantly, by extending the asymptotic expansion to first and second order in $\langle \text{Kn} \rangle$, we derive the governing ``continuum-level'' equation {\it and boundary conditions} for finite but small values of the Knudsen number ($\langle \text{Kn} \rangle \ll 1$). Specifically, for steady problems, the governing equation is shown to be the steady heat conduction equation up to order $\langle \text{Kn} \rangle ^2$,  while the corresponding boundary conditions are shown to be of the temperature-jump type, with jump coefficients that, in general, depend on the material and boundary properties. For unsteady problems, we show that the governing equation is the unsteady heat conduction equation up to first order in $\langle \text{Kn} \rangle$ with boundary conditions remaining the same as in the steady case up to that order for the case of prescribed-temperature boundaries. 

Jump boundary conditions have been observed before in solutions of the Boltzmann equation \cite{chen, lacroix05} and attempts were made \cite{chen} to explain these invoking differences in local equilibrium conditions across interfaces. The present work shows how temperature jumps arise as a result of the incompatibility between the isotropic distributions associated with boundary conditions and the anisotropic distribution associated with non-equilibrium resulting from transport (temperature gradients). A well-known manifestation of this physical behavior are the temperature jumps associated with the Kapitza interface problem. In section \ref{kapitza} we show how our asymptotic approach can be used to calculate the interface conductance (and associated temperature jump) from first principles (at the kinetic level, that is, given the interface transmission and reflection coefficient).

The temperature jump relations derived in this work are manifestations of what is known in the kinetic theory community as ``slip'', which gives its name to the slip regime, $0<\langle \text{Kn} \rangle \lesssim 0.1$. It is generally known \cite{Sone2002, cerc} that in this regime the material constitutive law may still be used {\it unmodified} and kinetic effects are accounted for by modified boundary conditions. In the field of rarefied gas dynamics, Cercignani  \cite{Cercignani1964} and Sone with co-workers \cite{sone69, sone77} were the first to provide systematic asymptotic solutions up to second order in $\langle \text{Kn} \rangle$, demonstrating the possibility of using the traditional ``continuum'' fluid dynamics, albeit with modified boundary conditions, beyond the slip regime and into the early transition regime. The transition regime is typically defined by $0.1 \lesssim \langle \text{Kn} \rangle \lesssim 10$ and represents the regime in which transport transitions from diffusive ($\langle \text{Kn} \rangle \ll 1$) to ballistic ($\langle \text{Kn} \rangle \gg 1$). Discussions of the use of asymptotic solutions of the Boltzmann equation in rarefied gas dynamics can be found in \cite{aoki2001, Sone2002, Sone2007}. 

The practical implications of the present work are twofold:   
first, solution of the heat equation is significantly easier (analytically or numerically) compared to the Boltzmann equation, especially in the regime $\langle \text{Kn} \rangle \ll 1$ where the latter becomes stiff. In addition to ease of solution, centuries of  investment in continumm formulations such as the heat equation, either in the form of education, mathematical solution techniques or numerical solution software, make this by far the preferred approach. This can be easily seen from the considerable efforts expended in developing approximate "effective thermal conductivity" concepts that enable the use of Fourier's law in the transition regime. The present work provides {\it rigorous} methods for obtaining solutions {\it consistent with the Boltzmann equation} in the slip and early transition regime. Studies in rarefied gas dynamics show that, depending on the problem and the amount of error that can be tolerated, slip/jump formulations could be used up to $\langle \text{Kn} \rangle \approx 0.5$ and sometimes beyond \cite{pof2006}. Second, by using the asymptotic solution as a control in deviational Monte Carlo schemes, one can overcome the stiffness associated with the $\langle \text{Kn} \rangle \ll 1$ regime. This happens because \cite{Peraud2014Adjoint,Radtke2013} the asymptotic solution becomes increasingly more accurate as $\langle \text{Kn} \rangle \rightarrow 0$, thus requiring increasingly less computational resources to describe the deviation therefrom as this limit is approached. This yields computational methods that are able to efficiently simulate problems characterized by $\langle \text{Kn} \rangle \ll 0.1$ locally or globally, in contrast to traditional Boltzmann solution methods.

The present paper is organized as follows: in section \ref{background} we introduce the governing (Boltzmann) equation and the notation used in this paper; in section \ref{Asymptotic_analysis}, we present the asymptotic analysis leading to derivation of the governing equation in the bulk up to second order in the Knudsen number.  Associated boundary conditions and boundary layer corrections up to first order in the Knudsen number are derived in section \ref{Order1BL}. In section \ref{Order2BL} we present results obtained from extending the boundary layer analysis to second order in Knudsen number. In section \ref{summary} we summarize and discuss our results and provide example applications to one-dimensional and two-dimensional problems. In section \ref{time-dep} we discuss the applicability of the asymptotic theory and its results (governing equations, boundary conditions and corrective boundary layers) to time-dependent problems. In section \ref{kapitza} we show how the asymptotic theory can be used to calculate the Kapitza conductance (and temperature jump) associated with the interface between two materials. We conclude with some final remarks in section \ref{final}.

\section{Background}
\label{background}
We consider the Boltzmann equation for phonon transport in the relaxation time approximation
\begin{equation}
\frac{\partial f}{\partial t'} + \mathbf{V}_g \cdot \nabla_{\mathbf{x}^\prime} f = \frac{f^\text{loc}-f}{\tau(\omega,p,T)}
\end{equation}
where $f=f(\mathbf{x}^\prime,\omega,p,\mathbf{\Omega},t')$ is the occupation number of the phonon states, $\mathbf{x}^\prime$ the position vector in physical space, $\mathbf{V}_g(\omega,p)$ the group velocity, $\omega$  the phonon frequency, $p$ the phonon  polarization, $\mathbf{\Omega}$ the unit vector denoting phonon traveling direction, $T$ the temperature and $f^\text{loc}$  an equilibrium distribution at the ``pseudotemperature'' $T_\text{loc}$ defined by energy conservation considerations (refer for instance to \cite{hao09,chen} for details on the definition of $f^\text{loc}$).

In this work we primarily consider steady problems. Extension to time-dependent problems directly follows by extending the methodology presented here. Scaling analysis in section \ref{time-dep} shows that, assuming diffusive time scaling, time dependence may modify the results presented here at order $\langle \text{Kn} \rangle^2$. In other words, the results obtained for steady state in this paper may be applied directly to order $\langle \text{Kn} \rangle^0$ and $\langle \text{Kn} \rangle^1$ with very few modifications, explained in section \ref{time-dep}.

Assuming small deviations from equilibrium at temperature $T_\text{eq}$, the linearized steady-state Boltzmann equation reads
\begin{equation}
\mathbf{V}_g \cdot \nabla_{\mathbf{x}^\prime} f^\text{d} = \frac{\mathcal{L}(f^\text{d})-{f^\text{d}}}{\tau(\omega,p,T_{\text{eq}})}
\end{equation}
where $f^\text{d} = (f-f^\text{eq})$, with $f^\text{eq} = [\exp(\hbar \omega/k_\text{b} T_\text{eq})-1]^{-1}$.

By noting that $\mathcal{L}(f^\text{d}) = (T_{\text{loc}}-T_{\text{eq}}) df^{\text{eq}}/dT$ and writing energy conservation \cite{hao09} in the form
\begin{equation}
\int_{\omega', p'} \mathcal{L}(f^\text{d}) \frac{D \hbar \omega'}{\tau}  d \omega' = \int_{\omega',p',\boldsymbol{\Omega}'} \frac{\hbar \omega' f^\text{d}}{\tau} \frac{D}{4\pi}d^2\boldsymbol{\Omega}' d\omega'
\label{energy_conservation}
\end{equation}
where $D=D(\omega,p)$ denotes the density of states, we obtain the expression
\begin{equation}
\mathcal{L}(f^\text{d})=\frac{\int_{\omega',p',\boldsymbol{\Omega}'} \frac{\hbar \omega' f^\text{d}}{\tau} \frac{D}{4\pi}d^2\boldsymbol{\Omega}' d\omega' }{C_\tau} \frac{df^\text{eq}}{dT}
\end{equation}
Here, and in what follows, unless otherwise stated, $\tau=\tau(\omega,p,T_\text{eq})$.
In the above expression, 
\begin{equation}
C_\tau = \int_{\omega,p} \frac{D\hbar \omega}{\tau}\frac{df^\text{eq}}{dT} d\omega
\end{equation}
Also, $\boldsymbol{\Omega}$ and $d^2\boldsymbol{\Omega}$ respectively refer to the unit vector defining the direction of propagation and to the differential solid angle, expressed as $\sin(\theta)d\theta d\phi$ in spherical coordinates.
In the interest of simplicity, in the above expressions and in what follows, we use a single integral symbol to denote both integrals over multiple variables and sum over polarization. 

In this study, relaxation times and group velocities may depend on frequency and polarization. For this reason, the Knudsen number is defined in an average sense. We choose the following (somewhat arbitrary) definition
\begin{equation}
\langle \text{Kn} \rangle = \frac{\int_{\omega,p} C_{\omega,p} \text{Kn}_{\omega,p}d\omega}{\int_{\omega,p}C_{\omega,p}d \omega}
\label{Kn_definition}
\end{equation}
 where
\begin{equation}
C_{\omega,p} = \hbar \omega D \frac{df^\text{eq}}{dT}
\end{equation}
and $\text{Kn}_{\omega,p} = \Lambda_{\omega,p}/L = V_g(\omega,p)\tau(\omega,p,T_\text{eq})/L$, which we will denote by Kn. In the expression for $\text{Kn}_{\omega,p}$, $V_g(\omega,p)=||\mathbf{V}_g(\omega,p)||$ is the magnitude of the group velocity

\section{Asymptotic analysis for the bulk}
\label{Asymptotic_analysis}
Introducing the dimensionless coordinate $\mathbf{x}=\mathbf{x}'/L$ as well as the normalization
\begin{equation}
\Phi = \frac{f^\text{d}}{\frac{df^\text{eq}}{dT}}
\label{definition_phi}
\end{equation}
we write the Boltzmann equation in the form
\begin{equation}
\boldsymbol{\Omega} \cdot \nabla_\mathbf{x} \Phi = \frac{\mathcal{L}(\Phi)-\Phi}{\text{Kn}}
\label{eq:dimensionless}
\end{equation}
where
\begin{equation}
\mathcal{L}(\Phi)=\frac{\int_{\omega,p,\boldsymbol{\Omega}} \frac{C_{\omega,p}}{ 4 \pi \tau} \Phi d^2\boldsymbol{\Omega} d\omega }{C_\tau}
\label{operator}
\end{equation}

The usual macroscopic quantities of interest such as  temperature, energy density and heat flux can be calculated from
\begin{align}
T_\text{tot}&=T_{\text{eq}} + \frac{1}{4\pi C} \int_{\omega,p,\boldsymbol{\Omega}} C_{\omega,p} \Phi d^2\Omega d\omega=T_\text{eq}+T(\mathbf{x}) \\
E_\text{tot}&=E_{\text{eq}} + \frac{1}{4\pi} \int_{\omega,p,\boldsymbol{\Omega}} C_{\omega,p} \Phi d^2\Omega d\omega \\
\mathbf{q}''&=\frac{1}{4 \pi}\int_{\omega,p,\boldsymbol{\Omega}} C_{\omega,p}V_g \Phi \boldsymbol{\Omega} d^2\Omega d \omega \label{qdef}
\end{align}
We will refer to $T(\mathbf{x})$ as the deviational temperature, since it represents the deviation from the equilibrium temperature $T_\text{eq}$.

\subsection{Bulk solution}

The asymptotic solution relies on a ``Hilbert-type'' \cite{grad63} expansion of the solution $\Phi$ in the form
\begin{equation}
\Phi=\sum_{n=0}^{\infty} \langle \text{Kn} \rangle^n \Phi_n
\label{phi_expansion}
\end{equation}
Given the nature of the proposed solution, similar expansions can be written  for the temperature and the heat flux fields
\begin{equation}
\begin{cases}
T&=\sum_{n=0}^{\infty} \langle \text{Kn}\rangle^n T_n \label{T_expansion}\\
\mathbf{q}''&=\sum_{n=0}^{\infty} \langle \text{Kn}\rangle^n \mathbf{q}_n''
\end{cases}
\end{equation}

In this section, we only consider solutions far from any boundary. As will be shown below, close to the boundary, kinetic effects become important due to the incompatibility of the bulk solution with the kinetic (Boltzmann) boundary condition and a separate, boundary layer analysis is required. 
Therefore, we let $\Phi_G=\sum \langle \text{Kn} \rangle^n\Phi_{Gn}$ be the bulk solution, anticipating that $\Phi=\Phi_G+\Phi_K$, where $\Phi_K$ represents kinetic boundary layer corrections that are zero in the bulk and will be similarly expanded later.
When the expansion for $\Phi_G$ is inserted in the Boltzmann equation we obtain
\begin{equation}
\boldsymbol{\Omega} \cdot \nabla_\mathbf{x} \sum_{n=0}^{\infty} \langle \text{Kn} \rangle^{n} \Phi_{Gn} = \sum_{n=0}^{\infty} \langle \text{Kn}\rangle^n \frac{\left[\mathcal{L}(\Phi_{Gn}) - \Phi_{Gn} \right]}{\text{Kn}}
\label{expanded_BTE}
\end{equation}
By equating terms of the same order ($\langle \text{Kn} \rangle^1$ and higher powers) and assuming that $\text{Kn}\sim \langle \text{Kn}\rangle$, we obtain the following relationship for all $n\geq 0$
\begin{equation}
\boldsymbol{\Omega} \cdot \nabla_\mathbf{x} \Phi_{Gn} =\frac{\langle \text{Kn} \rangle}{\text{Kn}} \left[ \mathcal{L}(\Phi_{Gn+1})-\Phi_{Gn+1} \right].
\label{eq:bulk_relationship}
\end{equation}
In addition, considering the two terms of order 0 in the right hand side of \eqref{expanded_BTE}, we find that $\Phi_{G0}$ is determined by the solution of the equation
\begin{equation}
\Phi_{G0}=\mathcal{L}(\Phi_{G0})=\frac{\int_{\omega,p,\boldsymbol{\Omega}}  \frac{C_{\omega,p}}{4 \pi \tau}\Phi_{G0} d^2\boldsymbol{\Omega} d\omega }{C_\tau}.
\label{phig0}
\end{equation}

The assumption $\text{Kn}\sim \langle \text{Kn}\rangle$ is easily satisfied when the range of free paths is relatively small (and is exactly satisfied in the single free path case $\Lambda_{\omega,p}=\Lambda=const$), but becomes harder to justify in materials with wide range of free paths. In the latter cases, it has the effect of reducing the value of $\langle \text{Kn} \rangle$ for which the theory presented here is valid. This is further discussed and quantified in section \ref{BL_results}.  

From equation (\ref{phig0}) we deduce that $\Phi_{G0}$ is a function that depends on $\mathbf{x}$ only, since this is the case for $\mathcal{L}(\Phi_{G0})$. We note here that any function that only depends on $\mathbf{x}$ is a solution. Additionally, since $\Phi_{G0}=\Phi_{G0}(\mathbf{x})$, we find that the zeroth order deviational bulk temperature is given by 
\begin{equation}
T_{G0}(\mathbf{x}) = \frac{1}{4\pi C} \int_{\omega,p,\boldsymbol{\Omega}} C_{\omega,p} \Phi_{G0}(\mathbf{x})d^2\boldsymbol{\Omega} d\omega =\Phi_{G0}(\mathbf{x}).
\end{equation}
and that
\begin{equation}
\mathbf{q}_{G0}'' = \frac{1}{4 \pi} \int_{\omega,p,\boldsymbol{\Omega}} C_{\omega,p} V_g \Phi_{G0}(\mathbf{x}) \boldsymbol{\Omega} d^2 \boldsymbol{\Omega} d\omega = 0
\end{equation}

At this stage, the spatial dependence of $\Phi_{G0}$ is undetermined. The additional information needed will be inferred from the application of a solvability condition to $\Phi_{G1}$. 
Using \eqref{eq:bulk_relationship} we find the following expression for the order 1 solution
\begin{equation}
\Phi_{G1}=\mathcal{L}(\Phi_{G1})-\frac{\text{Kn}}{\langle \text{Kn} \rangle }\boldsymbol{\Omega} \cdot \nabla_\mathbf{x} \Phi_{G0}
\label{phig1}
\end{equation}
This equation states that a necessary condition for  $\Phi_{G1}$ to be the order 1 solution is that it is equal to the sum of $-\text{Kn} \langle \text{Kn} \rangle^{-1}\boldsymbol{\Omega} \cdot \nabla_\mathbf{x} \Phi_{G0}$ and a function that only depends on $\mathbf{x}$.  Since the temperature associated with $\boldsymbol{\Omega} \cdot \nabla_\mathbf{x} \Phi_{G0}$ is zero, we can write 
\begin{equation}
\Phi_{G1}=T_{G1}-\text{Kn} \langle \text{Kn} \rangle^{-1} \boldsymbol{\Omega} \cdot \nabla_\mathbf{x} T_{G0}
\label{phig1v2}
\end{equation}

Finally, order 2 may be derived following the same procedure for eq (\ref{eq:bulk_relationship}) for $n=1$, which yields
\begin{equation}
\Phi_{G2} = \mathcal{L}(\Phi_{G2})-\frac{\text{Kn}}{\langle \text{Kn} \rangle } \boldsymbol{\Omega}\cdot \nabla_\mathbf{x} T_{G1} + \frac{\text{Kn}^2}{\langle \text{Kn} \rangle^2 }\boldsymbol{\Omega} \cdot \nabla_\mathbf{x} \left( \boldsymbol{\Omega} \cdot \nabla_\mathbf{x}T_{G0} \right)
\label{order_2_bulk}
\end{equation}
In the following section, while deriving the governing equation for $T_{G0}$, we also show that the temperature associated with $\Phi_{G2}$ is $\mathcal{L}(\Phi_{G2}) = T_{G2}$. 

\subsection{Governing equation for the temperature field}
\label{governing_equations_bulk}
The solvability condition required to determine $\Phi_{Gn}$ is the statement of energy conservation \eqref{energy_conservation} which, applied to $\Phi_{Gn+1}$, becomes
\begin{equation}
\int_{\omega,p} \frac{C_{\omega,p}}{\tau} \mathcal{L}(\Phi)d\omega = \int_{\omega,p,\boldsymbol{\Omega}} \frac{C_{\omega,p}}{4 \pi \tau} \Phi d\omega d^2\boldsymbol{\Omega}
\end{equation}
Using \eqref{eq:bulk_relationship} results in the following condition 
\begin{equation}
\int_{\omega,p,\boldsymbol{\Omega}}C_{\omega,p} V_{g} \boldsymbol{\Omega} \cdot \nabla_\mathbf{x} \Phi_{Gn} d\omega d^2\boldsymbol{\Omega}=0
\label{eq:solvability}
\end{equation}
that needs to be satisfied for all $n\geq 0$.
Applying this relationship to $\Phi_{G1}$, we obtain
\begin{equation}
\int_{\omega,p,\boldsymbol{\Omega}}C_{\omega,p} V_{g} \boldsymbol{\Omega} \cdot \nabla_\mathbf{x} \left( T_{G1}-\frac{\text{Kn}}{\langle \text{Kn} \rangle}\boldsymbol{\Omega} \cdot \nabla_\mathbf{x} T_{G0} \right) d\omega d^2\boldsymbol{\Omega}=0.
\end{equation}
In the above expression, the integral over the solid angle is zero in all terms where a component of the traveling direction appears with an odd exponent. This implies
\begin{equation}
\nabla_\mathbf{x}^2 T_{G0} = 0
\end{equation}
This concludes the proof that the 0-th order temperature field obeys the steady state heat equation. Moreover, from \eqref{order_2_bulk} it follows that 
\begin{equation}
\Phi_{G2} = T_{G2}-\frac{\text{Kn}}{\langle \text{Kn} \rangle } \boldsymbol{\Omega}\cdot \nabla_\mathbf{x} T_{G1} + \frac{\text{Kn}^2}{\langle \text{Kn} \rangle^2 }\boldsymbol{\Omega} \cdot \nabla_\mathbf{x} \left( \boldsymbol{\Omega} \cdot \nabla_\mathbf{x}T_{G0} \right)
\label{order2_bulk}
\end{equation}
In Appendix \ref{laplace_demo} we show that higher-order (in fact, possibly all order) terms similarly obey the heat equation. In other words, $T_{G1}(\mathbf{x})$ and $T_{G2}(\mathbf{x})$, are determined by solution of 
\begin{equation}
\nabla_\mathbf{x}^2 T_{G1} =0, \;\;\;\; \nabla_\mathbf{x}^2 T_{G2} =  0
\end{equation}

Before we close this section, we note that although in the Laplace-type equations derived above for the temperature the thermal conductivity does not appear, the above asymptotic analysis still clearly predicts that {\it in the bulk, the material constitutive relation (thermal conductivity) is equal to the "traditional" bulk value}. This can be seen from first-principles by inserting \eqref{phig1v2} into \eqref{qdef} to obtain
\begin{equation}
\langle \text{Kn} \rangle \mathbf{q}_{G1}''=-\frac{1}{4 \pi} \int_{\omega,p,\boldsymbol{\Omega}} \frac{V_g^2 \tau}{L} C_{\omega,p} \boldsymbol{\Omega} \left(\boldsymbol{\Omega} \cdot \nabla_\mathbf{x} T_0 \right) d\omega d^2 \boldsymbol{\Omega}=-\kappa \nabla_{\mathbf{x}'} T_{G0}
\end{equation}   
where the second equality follows from recognizing the well known expression 
\begin{equation}
\kappa=\frac{1}{3}\int_{\omega,p} V_g^2 \tau C_{\omega,p} d \omega  
\end{equation}

\section{Order 1 boundary layer analysis}
\label{Order1BL}
In this section, we extend the asymptotic analysis of the previous section to the vicinity of boundaries, where as will be shown below, a boundary layer analysis is required for matching the bulk solution of the previous section to the kinetic (BTE) boundary conditions of interest. Here we will consider two kinetic boundary conditions, namely, those of prescribed temperature and diffuse adiabatic reflection. In this work we assume that boundaries are flat; boundary curvature will be considered in a future publication. Without loss of generality we assume that the boundary is located at $x_1=0$ and with an inward normal pointing in the positive $x_1$ direction; $x_2$ and $x_3$ will denote cartesian coordinates in the plane of the boundary. Moreover, we will use $\Omega_1$, $\Omega_2$ and $\Omega_3$ to refer to the components of the unit vector $\boldsymbol{\Omega}$ in the coordinate system $(x_1,x_2,x_3)$.  In other words, $\Omega_1 = \cos(\theta)$, $\Omega_2 = \sin(\theta) \cos(\phi)$ and $\Omega_3 = \sin(\theta) \sin(\phi)$.

We now derive the general equation governing the boundary layer correction required in the boundary vicinity for matching the bulk solution to the kinetic (BTE) boundary conditions. We introduce the boundary layer function $\Phi_{K}$, written as a Hilbert expansion ($\Phi_n=\Phi_{Gn}+\Phi_{Kn}$) with $\Phi_{K0} = 0$ and insert it in the Boltzmann equation, obtaining
\begin{equation}
\sum_{i=1}^\infty \sum_{j=1}^3 \langle \text{Kn} \rangle^i \Omega_j \frac{\partial \Phi_{Ki}}{\partial x_j} =\sum_{i=1}^\infty \langle \text{Kn} \rangle^i \frac{\mathcal{L}(\Phi_{Ki}) - \Phi_{Ki}}{\text{Kn}}
\label{phi_K_equation}
\end{equation}

In the vicinity of the boundary, a new characteristic lengthscale, namely the distance from the boundary, becomes important. Similarly to \cite{Sone2002}, we introduce a ``stretched'' variable defined by $\eta = x_1/\langle \text{Kn} \rangle $. Equation \eqref{phi_K_equation} can thus be written in the form
\begin{equation}
\sum_{i=1}^\infty \langle \text{Kn} \rangle^{i-1} \Omega_1 \frac{\partial \Phi_{Ki}}{\partial \eta} =\sum_{i=1}^\infty \langle \text{Kn} \rangle^i \frac{\mathcal{L}(\Phi_{Ki}) - \Phi_{Ki}}{\text{Kn}} - \sum_{i=1}^\infty \langle \text{Kn} \rangle^{i} \left( \Omega_2 \frac{\partial \Phi_{Ki}}{\partial x_2}+\Omega_3 \frac{\partial \Phi_{Ki}}{\partial x_3} \right)
\end{equation}
By equating terms of the same order, we find that each boundary layer term is solution to a 1D (in physical space) Boltzmann-type equation. For $\Phi_{K1}$, this equation is
\begin{equation}
\Omega_1 \frac{\partial \Phi_{K1}}{\partial \eta} = \langle \text{Kn} \rangle \frac{\mathcal{L}(\Phi_{K1}) - \Phi_{K1}}{\text{Kn}}.
\label{eq:K1_boltz}
\end{equation}
The equations for $\Phi_{Kn}$, $n\geq 2$ include "volumetric source" terms resulting from the derivatives of the lower order boundary layers in the boundary tangential directions ($x_2$ and $x_3$). Specifically, for each order $i\geq 2$:
\begin{equation}
\Omega_1 \frac{\partial \Phi_{Ki}}{\partial \eta} = \langle \text{Kn} \rangle \frac{\mathcal{L}(\Phi_{Ki}) - \Phi_{Ki}}{\text{Kn}} - \left( \Omega_2 \frac{\partial \Phi_{K{i-1}}}{\partial x_2}+\Omega_3 \frac{\partial \Phi_{K{i-1}}}{\partial x_3} \right)
\label{higher_order_BLequation}
\end{equation}
The case $i=2$ will be considered in the following section, where second-order boundary layer analysis is carried out.

\subsection{Boundary conditions for prescribed temperature boundaries}
\label{boundary_conditions}
The term ``prescribed temperature boundary'' is typically used to describe a boundary approximating a black-body, absorbing incoming phonons and emitting phonons from an equilibrium (isotropic) distribution at a given temperature. In other words, the Boltzmann boundary condition associated with such a boundary at deviational temperature $T_b$ is a Bose-Einstein (equilibrium) distribution at the wall temperature, denoted here by $f^\text{eq}(\omega;T_\text{eq}+T_b)$. In the linearized case, the incoming distribution of deviational particles is therefore 
\begin{equation}
f_b=T_b\frac{df^\text{eq}}{dT}
\end{equation}
or simply, in terms of quantity $\Phi$ defined in \eqref{definition_phi}, $\Phi_b=T_b$.

We note that $\Phi_{G0}$ is isotropic and is thus able to match  $\Phi_b$ provided we set $T_{G0}= T_b$
at the boundary. Therefore, at order 0, the solution to the Boltzmann equation with prescribed temperature boundaries is given by the heat equation complemented by the traditional Dirichlet boundary conditions and no boundary layer correction is required ($\Phi_{K0}=0$, which also implies that $T_0=T_{G0}$).

This situation changes at order 1. The order 1 distribution $\Phi_{G1}=T_{G1}-\text{Kn} \langle \text{Kn} \rangle^{-1} \boldsymbol{\Omega} \cdot \nabla_\mathbf{x} T_{G0}$ is not isotropic due to the gradient of $T_{G0}$. As a consequence, there is a mismatch between the order 1 solution and the  boundary condition (which has been satisfied by $\Phi_{G0}$ and is thus zero for all subsequent orders). This mismatch can be corrected by introducing a boundary layer term $\Phi_{K1}$ governed by equation (\ref{eq:K1_boltz}) and subject to boundary condition $\Phi_{K1}|_{\eta=0}+\Phi_{G1}|_{\eta=0} = 0$,
which translates into the following relation
\begin{equation}
\Phi_{K1}|_{\eta=0} = -T_{G1}|_{\eta=0} + \frac{\text{Kn}}{\langle \text{Kn} \rangle}\boldsymbol{\Omega}\cdot \nabla_\mathbf{x} T_{G0}|_{\eta=0}
\label{BL_Boundary_condition}
\end{equation}

The term $\nabla_\mathbf{x} T_{G0}|_{\eta=0}$ is known from the order 0 solution. The term $T_{G1}|_{\eta=0}$ is unknown and determined by the fact that there exists only one value for $T_{G1}|_{\eta=0}$ such that $\Phi_{K1}$ tends to 0 for $\eta \rightarrow \infty$ \cite{Sone2002}. This determination proceeds by writing 
$ \Phi_{K1} = \Phi_{K1,1}+\Phi_{K1,2}+\Phi_{K1,3}$
where each of $\Phi_{K1,i},\ i=1, 2, 3$ is the solution to an equation of the form \eqref{eq:K1_boltz} with the associated  boundary condition:
\begin{align}
\Phi_{K1,i}|_{\eta=0}&=\left(-c_i+ \frac{\text{Kn}}{\langle \text{Kn} \rangle}\Omega_i \right) \left.  \frac{\partial T_{G0}}{\partial x_i}\right|_{\eta=0} \label{K1i}
\end{align}
Anticipating the values of $\Phi_{K1,i}$ to scale with $\partial T_{G0}/\partial x_i|_{\eta=0}$ in the above equations we have set $T_{G1}|_{\eta=0}=\sum_i c_i (\partial T_{G0}/\partial x_i)|_{\eta=0}$.
The constants $c_1,\,c_2,\,c_3$ are uniquely determined by the condition that $\Phi_{K1,1}$, $\Phi_{K1,2}$ and $\Phi_{K1,3}$ individually tend to zero for $\eta \rightarrow \infty$. 

One can easily verify that  for $i=2,3$, $c_i=0$, with 
\begin{equation}
\Phi_{K1,i} \equiv \Psi_{K1,i} \frac{\partial T_{G0}}{\partial x_i} = \begin{cases} \frac{\text{Kn}}{\langle \text{Kn} \rangle} \Omega_i  \left.\frac{\partial T_{G0}}{\partial x_i}\right|_{\eta=0}\exp \left( -\frac{\langle \text{Kn} \rangle \eta }{\text{Kn} \Omega_1 } \right), & \text{for } \Omega_1 >0 \\ 0, & \text{for } \Omega_1<0 \end{cases}
\label{expression_K1i}
\end{equation}
is a solution to \eqref{eq:K1_boltz} with boundary condition \eqref{K1i}. The temperature field associated with these functions is zero. Here we note that the above solutions have the property $\mathcal{L}(\Phi_{K1,2})=\mathcal{L}(\Phi_{K1,3})=0$ and thus are also solutions of  
\eqref{eq:K1_boltz} with the term $\mathcal{L}(\Phi_{K1})$ removed. We will use this observation throughout this paper for obtaining analytical solutions to a number of boundary layer problems.

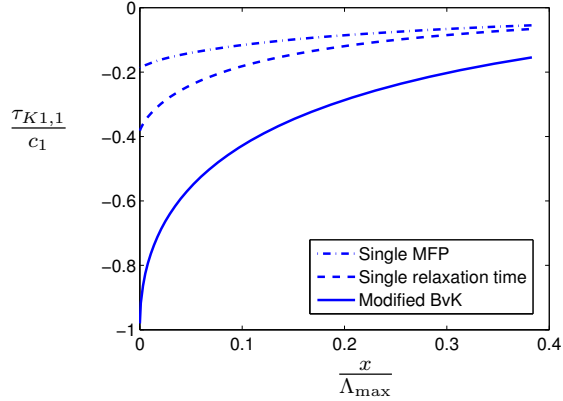
\begin{figure}[htbp]
\centering
	\includegraphics[width=.48\textwidth]{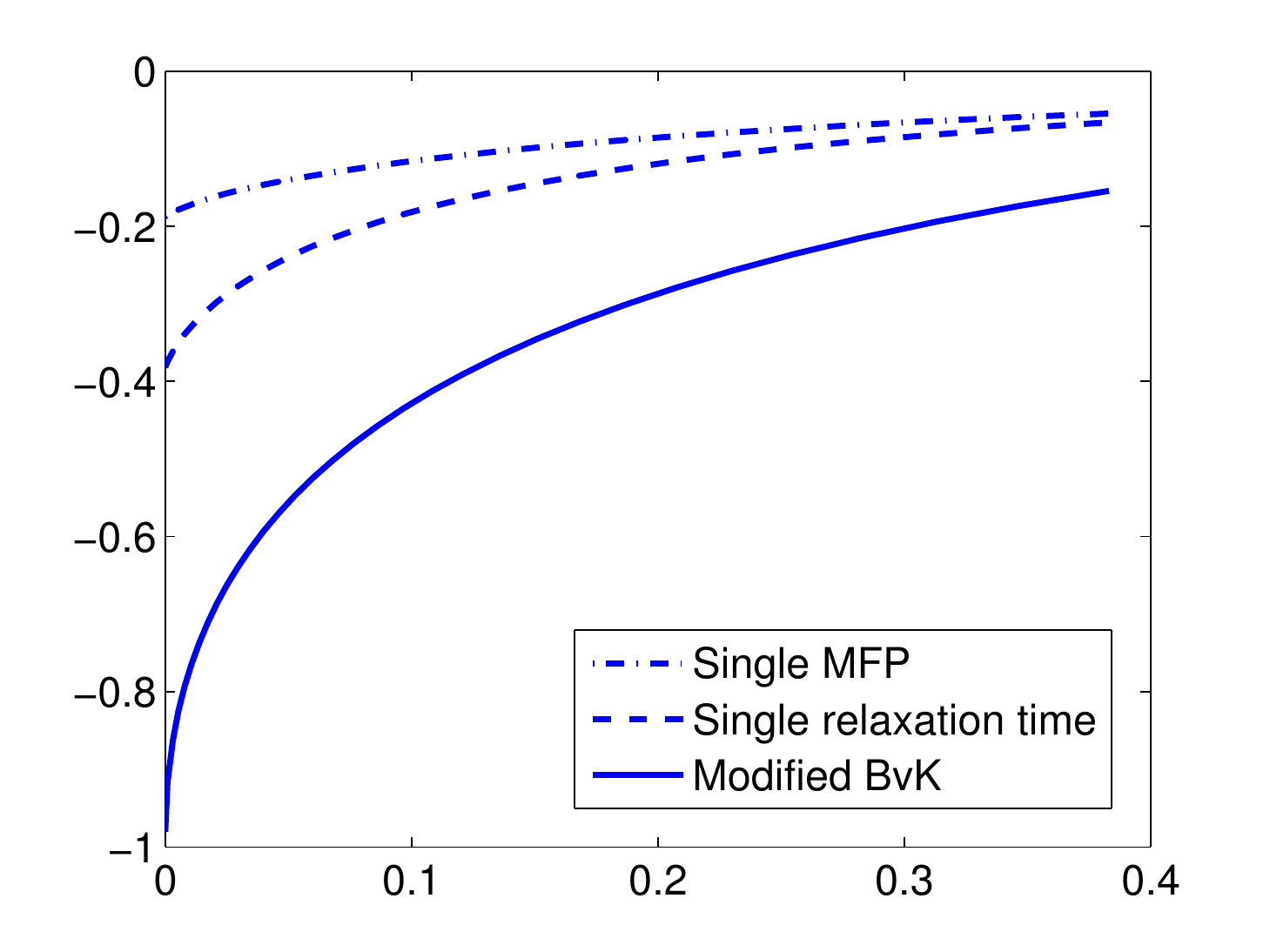}
	\caption{Temperature profile associated with $\tau_{K1,1} = T_{K1,1}/(\partial T_{G0}/\partial x_1)|_{\eta=0}$, for three relaxation time models. The $x$-axis is scaled by the maximum free path $\Lambda_{\text{max}}$ of each model.}
	\begin{tikzpicture}[overlay]
        		\draw[](1,2.2) node[left]{\rotatebox{0}{ \large $\frac{x}{\Lambda_{\text{max}}}$}};
		\draw[](-3.2,5.5) node[left]{\rotatebox{0}{ \large $\frac{\tau_{K1,1}}{c_1}$ }};

	\end{tikzpicture}
	\label{fig:order1_SMFP}
\end{figure}

The problem for $\Phi_{K1,1}$ must be solved numerically. Given the boundary condition it needs to satisfy, we write $\Phi_{K1,1}=\Psi_{K1,1}\left. (\partial T_{G0}/\partial x_1) \right|_{\eta=0}$ and solve for $\Psi_{K1,1}$. The numerical method developed and used for this purpose is explained in detail in Ref \cite{JPThesis}. In the case of a Debye and gray material  referred to here as the single free path case ($\text{Kn}=\langle \text{Kn} \rangle$ for all $\omega,p$), it yields $c_1= 0.7104$, while the resulting $\tau_{K1,1}\equiv \int_{\omega,p,\boldsymbol{\Omega}} C_{\omega,p} \Psi_{K1,1} d \omega d^2 \boldsymbol{\Omega} /4 \pi $ is plotted in Figure \ref{fig:order1_SMFP}. We note that Refs. \cite{Coron_1990,Li_2014} also report the value 0.7104 in the context of other kinetic particle transport, and develop other efficient methods for solving this problem.

In summary, the boundary condition for the order 1 bulk temperature field is
\begin{equation}
T_{G1}(x_1=0)=c_1\left. \frac{\partial T_{G0}}{\partial x_1}\right|_{x_1=0}
\end{equation}
or more generally 
\begin{equation}
T_{G1}\vert_{\mathbf{x}_b}=c_1\left. \frac{\partial T_{G0}}{\partial n}\right|_{\mathbf{x}_b}
\label{BC1}
\end{equation}
where  $\partial T_{G0}/\partial n$ refers to the derivative in the direction of the normal  to the boundary pointing into the material, $\mathbf{n}$, and $\mathbf{x}_b$ the boundary location.
In other words, the boundary condition is of the jump type and the associated temperature jump is proportional to the derivative of the 0th order solution in the direction normal to the boundary.

The amplitude of the corrective boundary layer that is added near the wall is also proportional to the normal derivative:
\begin{equation}
T_{K1,1} = \tau_{K1,1}\left. \frac{\partial T_{G0}}{\partial n}\right|_{\mathbf{x}_b}
\label{BL1}
\end{equation}
Note that although a non-zero temperature field is associated with $\Phi_{K1,1}$, the corresponding heat flux is zero. This is explained by the fact that $\Phi_{K1,1}$, by construction, tends to 0 at infinity. Since the boundary layer problem is one-dimensional in space, by energy conservation, the heat flux has to be constant in $x_1$ and is therefore zero everywhere. We also note that although $\Phi_{K1,2}$ and $\Phi_{K1,3}$ do not contribute to the temperature field, they do contribute in the heat flux $\mathbf{q}_{K1}''$ in the direction parallel to the boundary. Their contribution can be obtained by substituting \eqref{expression_K1i} into \eqref{qdef}; the result is summarized in section \ref{summary}.

\subsubsection{Numerical solution for complex material models}
\label{BL_results}
In section \ref{boundary_conditions} we reported the value of the coefficient $c_1$ and boundary-layer function $\Phi_{K1,1}$ in the single  free path case. In this section we report results for two more realistic material models. Specifically, we consider a material with realistic dispersion relation and a single relaxation time, as well as a material with realistic dispersion relation and frequency-dependent relaxation times. The dispersion relation in both cases is taken to be that of the [100] direction in silicon. The single relaxation time is taken to be 40ps. In the case of a variable relaxation time we use a slightly modified Born-von Karman-Slack (mBvKS) model \cite{dames2013} with parameters from \cite{minnichthesis} and \cite{Peraud2014Adjoint}, where the grain size used for boundary scattering is 0.27 mm instead of 2.7 mm. The reason for this approximation is that it facilitates the verification of the order 1 behavior with Monte Carlo simulation. We do not consider optical phonons in this work, but the method can be straightforwardly extended to this case.

We find $c_1=1.13$ in the single relaxation time model and $c_1=32.4$ in the mBvKS model. The associated boundary layers are plotted in figure \ref{fig:order1_SMFP}. It is important to note that:
\begin{itemize}
\item[-] The values of coefficient $c_1$ and the function $\tau_{K1,1}$ depends on the  definition of $\langle \text{Kn} \rangle$ or, equivalently, $\langle \Lambda \rangle$, which is rather arbitrary. This, however, does not influence the final result because the asymptotic temperature field, ultimately (see (\ref{T_expansion})) depends on the products $c_1 \langle \text{Kn} \rangle$ and $\tau_{K1,1}\langle \text{Kn} \rangle$  (see for instance solution \eqref{infinite_order}).
\item[-] The boundary layer in the mBvKS model is particularly wide (on the order of millimeters). This observation, as well as the large value of $c_1$, is a manifestation of the stiffness (multiscale nature) of this problem, resulting from the wide range of free paths present in this material;  mathematically, it is due to the factor $\text{Kn}/\langle \text{Kn} \rangle$ that appears in \eqref{BL_Boundary_condition} and which tends to give more weight to modes with very large free paths and makes the assumption $\text{Kn}\sim \langle \text{Kn}\rangle$ hard to satisfy. Since, by assumption, the sum of all $\Phi_{Gn} \langle \text{Kn} \rangle^n$ should exist --which requires $\Phi_{n} \langle \text{Kn} \rangle^n \lesssim 1$-- this has the overall effect of limiting the range of applicability of the asymptotic model to Knudsen numbers that are lower than the nominal $\langle \text{Kn} \rangle \lesssim 0.1$. It is important to note, however, that this limitation is a result of the fundamental physics of the problem: even at ``low'' Knudsen numbers given by $\langle \text{Kn}\rangle < 1/c_1$, there exist modes with long free paths (i.e. $\text{Kn}\sim O(0.1)$) introducing kinetic effects and making the zeroth order solution ($\nabla_\mathbf{x}^2 T_{G0}=0$) inadequate. 
\end{itemize}

\subsubsection{Validation}

\begin{figure}[htbp]
\centering
	\includegraphics[width=.48\textwidth]{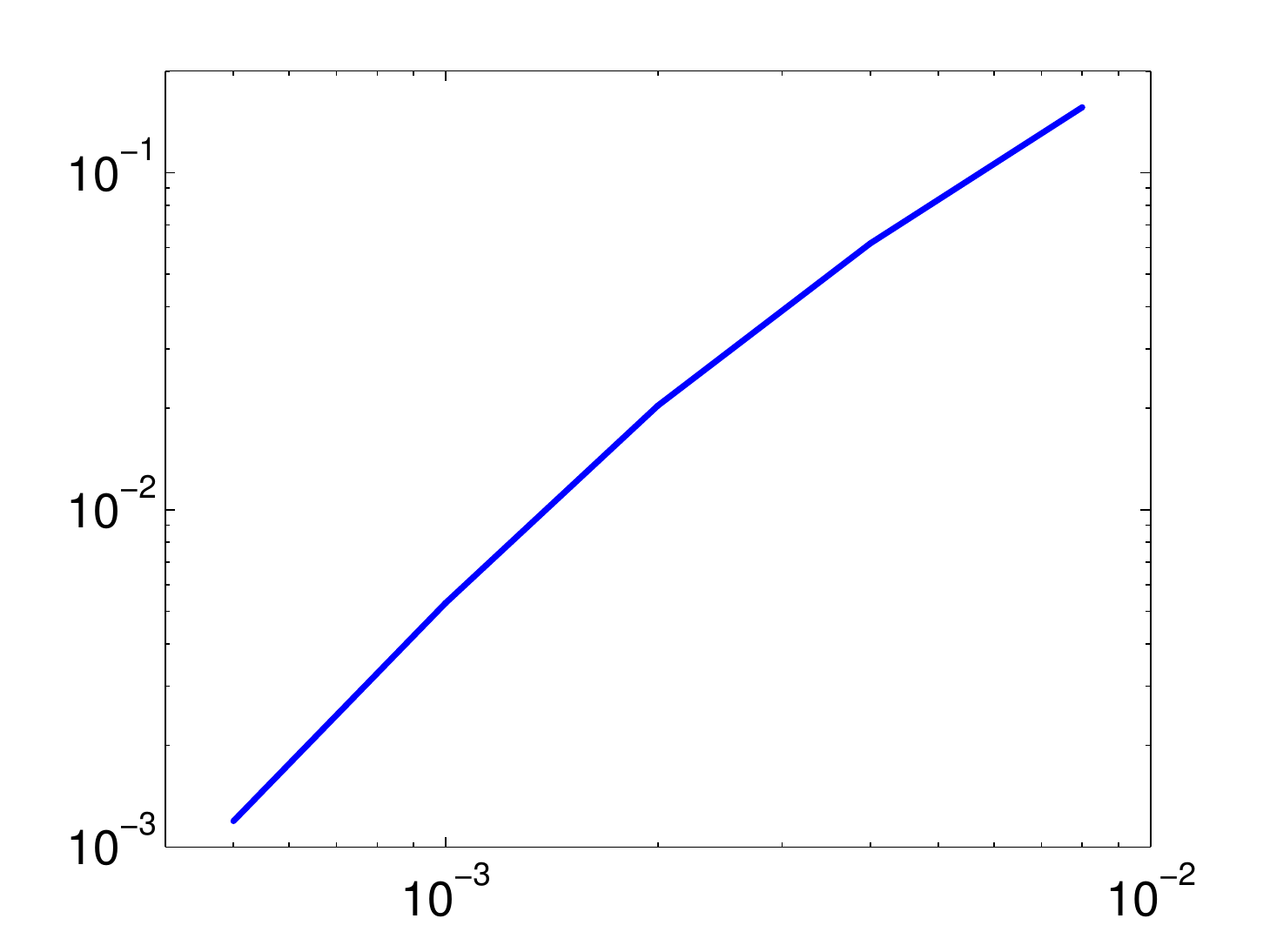}
	\caption{Validation of the first order asymptotic theory for prescribed temperature boundaries. The solid line denotes the normalized (by the 0th order, traditional Fourier, result) difference between the heat flux predicted by the asymptotic theory and MC simulation results. The dashed line denotes a slope of 2.}
	\begin{tikzpicture}[overlay]
        		\draw[](1,2.7) node[left]{\rotatebox{0}{ \large $\langle \text{Kn} \rangle$ }};
            \draw[](-3,5.6) node[left]{\rotatebox{90}{\small  $\displaystyle$ $\epsilon_q$}};
	\end{tikzpicture}
	\label{fig:1D_prescribed}
\end{figure}
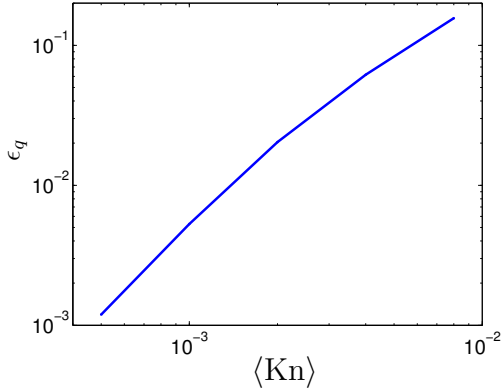

We validate our result using a one-dimensional problem, in which a mBvKS material is placed between two boundaries at prescribed temperatures and located at $x_1' = -L$ and $x_1'=L$, respectively. The order 0 (traditional Fourier) solution to this problem is a linear temperature profile $T_{G0}(x_1')$ which yields a heat flux $\kappa_{\text{Si-M}}\Delta T_{G0}/L$, where $\Delta T_{G0}$ is the temperature difference between the boundaries; here, $\kappa_{\text{Si-M}}$ denotes the bulk thermal conductivity associated with the mBvKS material. The temperature profile $T_{G1}$ is obtained by solving the Laplace equation with jump conditions
\begin{equation}
T_{G1}(x_1=\mp 1) = \pm c_1 \left. \frac{\partial T_{G0}}{\partial x_1}\right|_{x_1=\mp 1}
\end{equation}
and yields the modified heat flux $\kappa_{\text{Si-M}}(1-c_1 \langle \text{Kn}\rangle)\Delta T_{G0}/L $. We note that when calculated from an order $n$ temperature field, the heat flux is inherently an order $n+1$ quantity; in other words, the above result is correct to order 2. In Figure \ref{fig:1D_prescribed}, we plot the difference between the actual heat flux ($q_{x_1}''$, obtained using deviational Monte Carlo simulation \cite{peraud12,Peraud2014Adjoint}) and the asymptotic approximation, both normalized by $\kappa_{\text{Si-M}}\Delta T_{G0}/L $, namely, $\epsilon_q=q_{x_1}''L /(\kappa_{\text{Si-M}} \Delta T_{G0}) -(1-c_1 \langle \text{Kn} \rangle)$. The observed asymptotic behavior is order 2 which validates the order 1 accuracy of the asymptotic solution.

\subsection{Boundary condition for a diffuse adiabatic boundary}
\label{diffuse_order0}
The case of diffuse adiabatic boundaries can be treated through a similar approach, where the mismatch between the bulk asymptotic solution and the boundary condition is analyzed and corrected. The boundary condition at the kinetic level is given by \cite{mazumder01}
\begin{equation}
\left. \Phi\right|_{\mathbf{x}_b} = -\frac{1}{\pi}\int_{\Omega_1' <0}  \left. \Phi\right|_{\mathbf{x}_b} \Omega_1' d^2\mathbf{\Omega}' \text{ for } \Omega_1 >0
\label{diffuse}
\end{equation}
A major difference from the prescribed temperature boundary is that applying this condition to the 0th order bulk solution gives no information, because $\Phi_{G0}$ satisfies \eqref{diffuse} regardless of its value at the wall. The boundary condition for $T_{G0}$ is obtained by analyzing the order 1 mismatch. The order 1 boundary layer problem may be defined by applying the boundary condition \eqref{diffuse} to $\Phi_1=\Phi_{G1}+\Phi_{K1}$. It results in the following condition:
\begin{equation}
\begin{split}
&T_{G1}|_{\eta = 0}-\boldsymbol{\Omega} \cdot \nabla_\mathbf{x} T_{G0}|_{\eta = 0} + \Phi_{K1}|_{\eta = 0} = \\&-\frac{1}{\pi}\int_{\Omega_1'<0} \left( T_{G1}|_{\eta = 0}-\boldsymbol{\Omega}' \cdot \nabla_\mathbf{x} T_{G0}|_{\eta = 0} + \Phi_{K1}|_{\eta = 0} \right) \Omega_1'd^2\boldsymbol{\Omega}', \quad \quad \text{for } \Omega_1>0
\end{split}
\label{diffuse_o1_long}
\end{equation}
The isotropic term $T_{G1}$ readily cancels from both sides of the equality. Similarly to section \ref{boundary_conditions}, we define $\Phi_{K1} = \Phi_{K1,1}+\Phi_{K1,2}+\Phi_{K1,3}$ where each $\Phi_{K1,i}$ is associated with the temperature gradient in direction $i$  (as given by a right-handed set with $x_1$ being the direction normal to the boundary) and is a solution to the Boltzmann-type equation \eqref{eq:K1_boltz} with boundary condition:
\begin{equation}
-\left. \Omega_i \frac{\partial T_{G0}}{\partial x_i}\right|_{\eta = 0} + \Phi_{K1,i}|_{\eta = 0} = -\frac{1}{\pi}\int_{\Omega_1'<0} \left(-\Omega_i' \left. \frac{\partial T_{G0}}{\partial x_i}\right|_{\eta = 0} + \Phi_{K1,i}|_{\eta = 0} \right) \Omega_1' d^2\boldsymbol{\Omega}', \quad \quad \text{for }\Omega_1>0
\label{diffuse_o1_detailed}
\end{equation}
We find that solutions \eqref{expression_K1i}  satisfy the above conditions for $i=2$ and $i=3$ respectively, and do not impose any condition over the tangential derivatives of $T_{G0}$. For $i=1$, \eqref{diffuse_o1_detailed} results in
\begin{equation}
-\left( \frac{2}{3} + \Omega_1 \right) \left. \frac{\partial T_{G0}}{\partial x_1}\right|_{\eta = 0} =-\Phi_{K1,1}|_{\eta = 0} -2 \int_{\Omega_1'<0} \Phi_{K1,1}|_{\eta = 0} \Omega_1' d \Omega_1', \quad \quad \text{for }\Omega_1>0
\end{equation}
The only solution possible with this boundary condition is $\left. \Phi_{K1,1}\right|_{\eta=0}=\left.(\partial T_{G0}/\partial x_1)\right|_{\eta=0}=0$. This can be seen by noting that if $\left. (\partial T_{G0}/\partial x_1\right)|_{\eta=0}\neq 0$, multiplying the above equation by $\Omega_1$ and integrating over $0\leq \Omega_1 \leq 1$ yields 
$\int_{-1}^1 \Phi_{K1,1}|_{\eta=0} \Omega_1 d\Omega_1 \neq 0$, which is impossible (this can be seen by starting from the equation governing $\Phi_{K1,1}$--of the type \eqref{eq:K1_boltz}--and integrating over $0\leq \eta\leq \infty$ and $-1\leq \Omega_1\leq 1$ and using the condition $\Phi_{K1,1}(\eta\rightarrow \infty)\rightarrow 0$).
We thus conclude that $T_{G0}$ must satisfy the boundary condition
\begin{equation}
\left. \frac{\partial{T_{G0}}}{\partial n}\right|_{\mathbf{x}_b}
 =0,
\label{order0_diffuse}
\end{equation}
which is agrees with the Neumann boundary conditions associated with adiabatic boundaries.
\section{Order 2 boundary layer analysis}
\label{Order2BL}
\subsection{Order 2 analysis for prescribed temperature boundaries}
\label{Order2BL_prescribed}
The second order correction $\Phi_{K2}$ must be solution of \eqref{higher_order_BLequation} for $i=2$, namely 
\begin{equation}
\Omega_1 \frac{\partial \Phi_{K2}}{\partial \eta} = \langle \text{Kn} \rangle \frac{\mathcal{L}(\Phi_{K2}) - \Phi_{K2}}{\text{Kn}} - \left( \Omega_2 \frac{\partial \Phi_{K1}}{\partial x_2}+\Omega_3 \frac{\partial \Phi_{K1}}{\partial x_3} \right)
\label{second_order_BLequation}
\end{equation}
with the boundary conditions
\begin{multline}
\Phi_{K2}|_{\eta=0} = -\Phi_{G2}|_{\eta=0}=-T_{G2}|_{\eta=0} + \frac{\text{Kn}}{\langle \text{Kn} \rangle} \sum_i \Omega_i \left. \frac{\partial T_{G1}}{\partial x_i} \right|_{\eta=0} - \frac{\text{Kn}^2}{\langle \text{Kn} \rangle^2} \sum_{i,j}\Omega_i \Omega_j \left. \frac{\partial^2 T_{G0}}{\partial x_i \partial x_j} \right|_{\eta=0} \\ \quad \text{for } \Omega_1>0
\label{BC2_basic}
\end{multline}
Here we note that the derivatives of the first order boundary layer which appear in the RHS of \eqref{second_order_BLequation} introduce four volumetric source terms in the governing equation. 

The boundary condition \eqref{BC2_basic} includes three terms with first order partial derivatives of $T_{G1}$ and nine terms with second order derivatives. Taking into account the four source terms on the RHS of \eqref{second_order_BLequation}, we introduce sixteen constants such that the order 2 ``temperature jump'', $T_{G2}|_{\eta=0}$, may be written as
\begin{equation}
T_{G2}|_{\eta=0} = \sum_{i=1}^3 d_i \left. \frac{\partial T_{G1}}{\partial x_i}\right|_{\eta = 0}  +\sum_{i,j=1}^3 g_{ij} \left. \frac{\partial^2 T_{G0}}{\partial x_i \partial x_j}\right|_{\eta = 0}  + \sum_{i,j=2}^3\tilde{g}_{ij} \left. \frac{\partial^2 T_{G0}}{\partial x_i \partial x_j} \right|_{\eta = 0} .
\label{Order2Coefficients}
\end{equation}
We accordingly introduce sixteen boundary layer functions such that the total order 2 boundary layer may be written as:
\begin{equation}
\Phi_{K2} = \sum_{i=1}^3 \Psi_{K2,i} \left. \frac{\partial T_{G1}}{\partial x_i}\right|_{\eta = 0}  +\sum_{i,j=1}^3 \Psi_{K2,ij} \left. \frac{\partial^2 T_{G0}}{\partial x_i \partial x_j}\right|_{\eta = 0}  + \sum_{i,j=2}^3 \tilde{\Psi}_{K2,ij} \left. \frac{\partial^2 T_{G0}}{\partial x_i \partial x_j} \right|_{\eta = 0} 
\label{Order2Functions}
\end{equation}

The 16 unknown coefficients and boundary layer functions can be determined using a combination of numerical and analytical techniques; these are discussed in Appendix \ref{Second order prescribed}. Here we summarize the final result, which, conveniently, is quite compact. The second order temperature jump is given by the condition
\begin{equation}
T_{G2}|_{\eta=0}=c_1 \left. \frac{\partial T_{G1}}{\partial n}\right|_{\eta=0}.
\label{BC2}
\end{equation}
Due to its simplicity and compactness, this result lends itself particularly well to implicit application of boundary conditions; this is discussed in section \ref{implicit}.
The analogy to the order one temperature jump extends to the temperature boundary layer that is given by
\begin{equation}
T_{K2,1}=\tau_{K1,1} \left. \frac{\partial T_{G1}}{\partial n}\right|_{\eta=0}.
\end{equation}
In addition to this temperature boundary layer, the analysis yields a second order heat flux boundary layer. It may be calculated analytically by inserting expression \eqref{Order2Functions} for $\Phi_{K2}$  into
\begin{equation}
\mathbf{q}_{K2}''(\eta) = \int_{\omega,p,\boldsymbol{\Omega}}\frac{C_{\omega,p}}{4 \pi}\Phi_{K2} \mathbf{V}_g d^2 \boldsymbol{\Omega} d \omega,
\label{eq:HeatBL}
\end{equation}
which can be written in terms of incomplete Gamma functions. Validation of these results can be found in \cite{JPThesis}.

\subsection{Order 2 analysis of a diffusely reflective boundary}
\label{diffuse_order1}
In section \ref{diffuse_order0}, we resorted to an analysis of the order 1 boundary layers to obtain the order 0 boundary condition, and showed the latter amounts to the well-known Neumann boundary condition. Similarly, we here proceed with the  order 2 analysis in order to find the boundary condition for the order 1 temperature field.

Inserting \eqref{order2_bulk} in \eqref{diffuse} and introducing a boundary layer term yields, for $\Omega_1>0$ and for all frequency/polarization modes:
\begin{multline}
\label{diffuse_order_1equation}
T_{G2}|_{\mathbf{x}_b} - \frac{\text{Kn}}{\langle \text{Kn} \rangle} \boldsymbol{\Omega}\cdot \nabla_\mathbf{x} T_{G1}|_{\mathbf{x}_b} 
+ \frac{\text{Kn}^2}{\langle \text{Kn} \rangle^2} \boldsymbol{\Omega}\cdot \nabla_\mathbf{x} \left(\boldsymbol{\Omega} \cdot \nabla_\mathbf{x}  T_{G0}\right)|_{\mathbf{x}_b}
 +\Phi_{K2}|_{\mathbf{x}_b}
 =\\ -\frac{1}{\pi} \int_{\Omega_1'<0} \Omega_1' \left( T_{G2}|_{\mathbf{x}_b} 
 - \frac{\text{Kn}}{\langle \text{Kn} \rangle} \boldsymbol{\Omega'} \cdot \nabla_\mathbf{x} T_{G1}|_{\mathbf{x}_b} + 
 \frac{\text{Kn}^2}{\langle \text{Kn} \rangle^2} \boldsymbol{\Omega'}\cdot \nabla_\mathbf{x} \left(\boldsymbol{\Omega'} \cdot \nabla_\mathbf{x}  T_{G0}\right)|_{\mathbf{x}_b} +\Phi_{K2}|_{\mathbf{x}_b} \right) d^2 \boldsymbol{\Omega}'
\end{multline}
Moving to the coordinate system $(x_1,x_2,x_3)$ and the stretched coordinate $\eta$, we first note that in \eqref{diffuse_order_1equation}, the  derivatives
$\left. \partial^2 T_{G0}/(\partial x_i \partial x_1) \right|_{\eta=0}$
are zero for $i=2,3$ because $(\partial T_{G0}/\partial x_1)|_{\eta =0}=0$.

Boundary layer $\Phi_{K2}$ may be decomposed into 4 components, $\Phi_{K2,1}$, $\Phi_{K2,2}$, $\Phi_{K2,3}$ and $\Phi_{K2,23}$. Components $\Phi_{K2,2}$ and $\Phi_{K2,3}$ are similar to the order 1 boundary layers $\Phi_{K1,2}$ and $\Phi_{K1,3}$ (see expression \eqref{expression_K1i}), with the only difference being that $T_{G0}$ is replaced by $T_{G1}$. Component $\Phi_{K2,23}$ corrects the anisotropic mismatch associated with the bulk term $2\Omega_2 \Omega_3 \partial^2 T_{G0}/(\partial x_2 \partial x_3)$. It is a solution to the 1D Boltzmann equation \eqref{eq:K1_boltz} with boundary condition
\begin{equation}
\left. \Phi_{K2,23}\right|_{\eta=0}=-2 \Omega_2 \Omega_3 \left. \frac{\partial^2 T_{G0}}{\partial x_2 \partial x_3}\right|_{\eta=0}
\end{equation}
for $\Omega_1>0$, and 0 at infinity, and is therefore given by
\begin{equation}
\Phi_{K2,23}=-2 \Omega_2 \Omega_3 \left. \frac{\partial^2 T_{G0}}{\partial x_2 \partial x_3}\right|_{\eta=0} \exp \left( \frac{-\eta \langle \text{Kn} \rangle }{\Omega_1 \text{Kn} } \right) H(\Omega_1)
\end{equation}
Components $\Phi_{K2,2}$, $\Phi_{K2,3}$ and $\Phi_{K2,23}$ do not contribute to a temperature jump or (temperature) corrective layer, but they do contribute to the heat flux boundary layer.  

The last component is solution to the following problem:
\begin{equation} \label{BLO2_diffuse}
\left \{
\begin{split}
&\Omega_1 \frac{\partial \Phi_{K2,1}}{\partial \eta} = \frac{\langle \text{Kn} \rangle}{\text{Kn}} \left( \mathcal{L}(\Phi_{K2,1})-\Phi_{K2,1} \right)-\sum_{i=2}^3\frac{\text{Kn}}{\langle \text{Kn} \rangle} \Omega_i^2 \left. \frac{\partial^2 T_{G0}}{\partial x_i^2} \right|_{\eta = 0}  \exp \left( \frac{-\eta \langle \text{Kn} \rangle }{\Omega_1 \text{Kn} } \right) H(\Omega_1) \\
&\begin{split} -\frac{\text{Kn}}{\langle \text{Kn} \rangle} &\left( \frac{2}{3}+\Omega_1 \right) \left. \frac{\partial T_{G1}}{\partial x_1} \right|_{\eta = 0} + \frac{\text{Kn}^2}{\langle \text{Kn} \rangle^2} \left( \Omega_1^2 -\frac{1}{2} \right) \left. \frac{\partial^2 T_{G0}}{\partial x_1^2} \right|_{\eta = 0} + \frac{\text{Kn}^2}{\langle \text{Kn} \rangle^2} \sum_{i=2}^3 \left( \Omega_i^2 -\frac{1}{4} \right) \left. \frac{\partial^2 T_{G0}}{\partial x_i^2} \right|_{\eta = 0} 
\\&+\Phi_{K2,1}|_{\eta = 0}
 = -\frac{1}{\pi} \int_{\Omega_1'<0} \Omega_1' \Phi_{K2,1}|_{\eta = 0} d^2 \boldsymbol{\Omega}' , \quad \text{for} \ \Omega_1>0 \ \text{and all} \ \omega,p \end{split} \\
&\lim_{\eta \to\infty} \Phi_{K2,1}(\boldsymbol{\Omega},\omega,p,\eta) =0
\end{split}
\right.
\end{equation}

Although we could solve problem \eqref{BLO2_diffuse} using the numerical method  described in \cite{JPThesis}, we will here directly find the value of $\gamma$ without specifically calculating $\Phi_{K2,1}$. We first proceed by multiplying the boundary condition (second equation of problem \eqref{BLO2_diffuse}) by $\Omega_1$ and integrating over the half sphere described by $\Omega_1>0$ to obtain
\begin{equation}
\int_{\boldsymbol{\Omega}} \Omega_1 \Phi_{K2,1}|_{\eta = 0}d^2 \boldsymbol{\Omega} = \frac{4\pi}{3} \frac{\text{Kn}}{\langle \text{Kn} \rangle} \frac{\partial T_{G1}}{\partial x_1}
\end{equation}
We also multiply the first equation of problem \eqref{BLO2_diffuse} by $V_g C_{\omega,p}$ and integrate it over all frequencies and solid angles and $0\leq \eta < \infty$ to obtain
\begin{multline}
\label{BLO2_diffuse_integrated}
\left[ \int_{\boldsymbol{\Omega},\omega,p} C_{\omega,p}V_g \Omega_1 \Phi_{K2,1}|_{\eta \rightarrow \infty} d \omega d^2 \boldsymbol{\Omega} - \int_{\boldsymbol{\Omega},\omega,p} C_{\omega,p}V_g \Omega_1  \Phi_{K2,1}|_{\eta =0}d \omega d^2 \boldsymbol{\Omega} \right] \\= -\frac{\pi}{4} \int_{\omega,p} V_g \frac{\text{Kn}^2}{\langle \text{Kn} \rangle^2} C_{\omega,p} d \omega \sum_{i=2}^3 \frac{\partial^2 T_{G0}}{\partial x_i^2}
\end{multline}
Since $\Phi_{K2,1}$ tends to 0 at infinity and $\nabla_\mathbf{x}^2 T_{G0}=0$, we deduce the jump relation
\begin{equation}
\left. \frac{\partial T_{G1}}{\partial x_1}\right|_{\eta = 0} = \gamma \left. \frac{\partial^2 T_{G0}}{\partial x_1^2}\right|_{\eta = 0}
\label{diffuse_O1_condition}
\end{equation}
with
\begin{equation}
\gamma = -\frac{3}{16} \frac{\int_{\omega,p} \text{Kn}^2 V_g C_{\omega,p} d \omega}{\langle \text{Kn} \rangle \int_{\omega,p} \text{Kn} V_g C_{\omega,p} d \omega},
\end{equation}
which can be rewritten in the form
\begin{equation}
\gamma = -\frac{3}{16} \frac{\int_{\omega,p} V_g^3 \tau^2 C_{\omega,p} d \omega}{\langle \Lambda \rangle \int_{\omega,p} V_g^2 \tau C_{\omega,p} d \omega}.
\label{gamma_def}
\end{equation}
In the single free path model, $\gamma=-3/16$. Validation of this result can be found in \cite{JPThesis}. Note also that the approach that we used for finding $\gamma$ may be used for finding the heat flux associated with the boundary layer $\Phi_{K2,1}$. 

{\it A note on the physical interpretation of \eqref{diffuse_O1_condition}}
At first glance, the boundary condition \eqref{diffuse_O1_condition} seems to suggest that energy is not conserved since the net heat flux into the (diffusely reflective) boundary is not zero. In fact, contrary to appearances, this form  {\it ensures} energy conservation at the boundary. This can be seen by considering that $\partial^2 T_{G0}/\partial x_1^2\neq 0$ (only possible in two or three dimensions) implies variations in the temperature gradient {\it along} the boundary, which in turn implies variations in the heat flux along the boundary due to first-order kinetic boundary layers (see \eqref{expression_K1i}). Imposing energy conservation at the boundary reveals that \eqref{diffuse_O1_condition} exactly balances the terms resulting from gradients along the boundary \cite{JPThesis}. 

\section{Summary and discussion of results}
\label{summary}
 We have derived the continuum equations and associated boundary conditions that provide solutions equivalent to those of the Boltzmann equation up to second-order in Knudsen number for steady problems. This derivation shows that the governing equation in the bulk, up to at least second order in Knudsen number, is the steady heat conduction equation {\it with the bulk thermal conductivity}. Kinetic effects, always present at the boundaries due to the inhomogeneity introduced by the boundary and the concomitant mismatch between the distribution introduced by the kinetic (Boltzmann) boundary condition and the distribution function in the bulk, become increasingly important (can be observed in larger parts of the physical domain) as the Knudsen number increases. Fortunately, these kinetic effects can be systematically described and incorporated into the continuum solution relatively straightforwardly via the addition of kinetic boundary layer functions that are universal for a given material and material-boundary interaction model. 

We have studied two types of kinetic boundary conditions: prescribed wall temperature and diffuse reflection. We now summarize the procedure for obtaining the temperature and heat flux fields for an arbitrary problem of interest.

{\it Prescribed wall temperature boundary condition}:
Let $T_b(\mathbf{x}_b)$ denote the prescribed temperature along the system boundary denoted by $\mathbf{x}_b$ with boundary normal $\mathbf{n}$. According to the asymptotic theory, the temperature and heat flux fields can be calculated from
\begin{eqnarray}
{T(\mathbf{x})}&=&{T_0(\mathbf{x})}+\langle \text{Kn} \rangle(\,{T_{G1}(\mathbf{x})}+{T_{K1}(\mathbf{x})}\,)+ \langle \text{Kn}\rangle^2(\,{T_{G2}(\mathbf{x})}+{T_{K2}(\mathbf{x})}\,)+O(\langle \text{Kn} \rangle^3)\nonumber\\
{\mathbf{q}''(\mathbf{x})}&=&\langle \text{Kn} \rangle(\,{\mathbf{q}_{G1}''(\mathbf{x})}+{\mathbf{q}_{K1}''(\mathbf{x})}\,)+\langle \text{Kn}\rangle^2(\,{\mathbf{q}_{G2}''(\mathbf{x})}+{\mathbf{q}_{K2}''(\mathbf{x})}\,)+O(\langle \text{Kn}\rangle^3)\nonumber
\end{eqnarray}
where 
\begin{itemize}
\item $T_0(\mathbf{x})$ is solution to $\nabla_\mathbf{x}^2T_0=0$ subject to $T_0|_{\mathbf{x}_b}=T_b|_{\mathbf{x}_b}$
\item $T_{G1}(\mathbf{x})$ is solution to $\nabla_\mathbf{x}^2T_{G1}=0$ subject to $T_{G1}|_{\mathbf{x}_b}=c_1\frac{\partial T_{0}}{\partial \mathbf{n}}\vert_{\mathbf{x}_b}$
\item $T_{G2}(\mathbf{x})$ is solution to $\nabla_\mathbf{x}^2T_{G2}=0$ subject to $T_{G2}|_{\mathbf{x}_b}=c_1\frac{\partial T_{G1}}{\partial \mathbf{n}}\vert_{\mathbf{x}_b}$
\item $T_{K1}(\mathbf{x})=\tau_{K1,1}(\eta) \frac{\partial T_{0}}{\partial \mathbf{n}}\vert_{\mathbf{x}_b}$
\item $T_{K2}(\mathbf{x})=\tau_{K1,1}(\eta) \frac{\partial T_{G1}}{\partial \mathbf{n}}\vert_{\mathbf{x}_b}$
\item $\langle \text{Kn}\rangle \mathbf{q}_{Gi}=-\kappa \nabla_\mathbf{x'} T_{Gi-1}$, $i=1,2$
\item $\mathbf{q}_{K1}''(\mathbf{x})=\sum_{i=2}^3 \int_{\omega,p,\boldsymbol{\Omega}} \frac{C_{\omega,p} V_g}{4 \pi}  \Omega_i \Psi_{K1,i}(\eta)   d \omega d^2 \boldsymbol{\Omega} \left. \frac{\partial T_{0}}{\partial x_i} \right|_{\mathbf{x}_b} \mathbf{e}_i$ with $\Psi_{K1,i}, i=2,3$ given by \eqref{expression_K1i}.
\item $\mathbf{q}_{K2}''(\mathbf{x})=\int_{\omega,p,\boldsymbol{\Omega}} \frac{C_{\omega,p}}{4 \pi} \mathbf{V}_g \Phi_{K2}(\eta) d \omega d^2 \boldsymbol{\Omega}$  with $\Phi_{K2}$ given by \eqref{Order2Functions}.
\end{itemize}
We recall here that the coordinate $\eta$ is a stretched (by $\langle \text{Kn}\rangle^{-1}$) version of the local normal to the boundary. The boundary layer functions $\tau_{K1,1}(\eta)$, $\Psi_{K1,i}(\eta)$ and $\Psi_{K2}(\eta)$ are unique (universal) for each material and material-boundary interaction model. Figure \ref{fig:order1_SMFP} shows results for $\tau_{K1,1}(\eta)$ for three material models. The method for calculating this function is described in detail in \cite{JPThesis}. The boundary layer functions $\Psi_{K1,i}(\eta)$ and $\Psi_{K2}(\eta)$ are known analytically.
We also note that due to the absence of kinetic boundary layer corrections, at order zero $T_{G0}=T_0$.

{\it Diffusely reflecting boundary}:
In the case of a diffusely reflecting boundary located at $\mathbf{x}_b$ with normal vector $\mathbf{n}$, the temperature and heat flux fields can be calculated from
\begin{eqnarray}
{T(\mathbf{x})}&=&{T_0(\mathbf{x})}+\langle \text{Kn} \rangle {T_{G1}(\mathbf{x})}+O(\langle \text{Kn}\rangle^2)\nonumber\\
{\mathbf{q}''(\mathbf{x})}&=&\langle \text{Kn}\rangle ( {\mathbf{q}_{G1}''(\mathbf{x})}+{\mathbf{q}_{K1}''(\mathbf{x})}\,)+\langle \text{Kn}\rangle ^2({\mathbf{q}_{G2}''(\mathbf{x})}+{\mathbf{q}_{K2}''(\mathbf{x})}\,)+O(\langle \text{Kn}\rangle^3)\nonumber
\end{eqnarray}
where 
\begin{itemize}
\item $T_0(\mathbf{x})$ is solution to $\nabla_\mathbf{x}^2T_0=0$ subject to $\frac{\partial T_{0}}{\partial \mathbf{n}}\vert_{\mathbf{x}_b}=0$
\item $T_{G1}(\mathbf{x})$ is solution to $\nabla_\mathbf{x}^2T_{G1}=0$ subject to $\frac{\partial T_{G1}}{\partial \mathbf{n}}\vert_{\mathbf{x}_b}=\gamma \frac{\partial^2 T_{0}}{\partial \mathbf{n}^2}\vert_{\mathbf{x}_b}$ with $\gamma$ given by \eqref{gamma_def}.
\item $\langle \text{Kn}\rangle \mathbf{q}_{Gi}''=-\kappa \nabla_\mathbf{x'} T_{Gi-1}$, $i=1,2$
\item $\mathbf{q}_{K1}''(\mathbf{x})=\sum_{i=2}^3 \int_{\omega,p,\boldsymbol{\Omega}} \frac{C_{\omega,p} V_g}{4 \pi}  \Omega_i \Psi_{K1,i}   d \omega d^2 \boldsymbol{\Omega} \left. \frac{\partial T_{0}}{\partial x_i} \right|_{\mathbf{x}_b} \mathbf{e}_i$ with $\Psi_{K1,i}, i=2,3$ given by \eqref{expression_K1i}.
\item $\mathbf{q}_{K2}''(\mathbf{x})=\int_{\omega,p,\boldsymbol{\Omega}} \frac{C_{\omega,p}}{4 \pi} \mathbf{V}_g \Phi_{K2}(\eta) d \omega d^2 \boldsymbol{\Omega}$  with the components of $\Phi_{K2}$ given in section \ref{diffuse_order1}.
\end{itemize}
We note here that $\Psi_{K1,i}$ is identical to the corresponding boundary layer function that appeared in the prescribed-temperature boundary condition case. We also note that due to the structure of the boundary-layer problem for the diffusely reflecting boundary, the first-order analysis yields a zeroth order boundary condition, while a second-order analysis yields a first order boundary condition; as a result the asymptotic solution for the temperature terminates at first order in $\langle \text{Kn} \rangle$.

We see that, in both cases, the "traditional" Fourier description corresponds to the zeroth order solution.
\subsection{A one-dimensional example}
\label{couette-example}
In this section we consider a simple 1D problem as a means of illustrating the application of the asymptotic theory to problems of interest. We consider a silicon slab of thickness $L$ confined between two boundaries at different prescribed temperatures. Using dimensionless coordinates, the boundaries are located at $x_1=-1/2$ and $x_1=1/2$ and have deviational temperatures $T_L$ and $T_R$, respectively. 

\begin{figure}[htbp]
\centering
	\includegraphics[width=.68\textwidth]{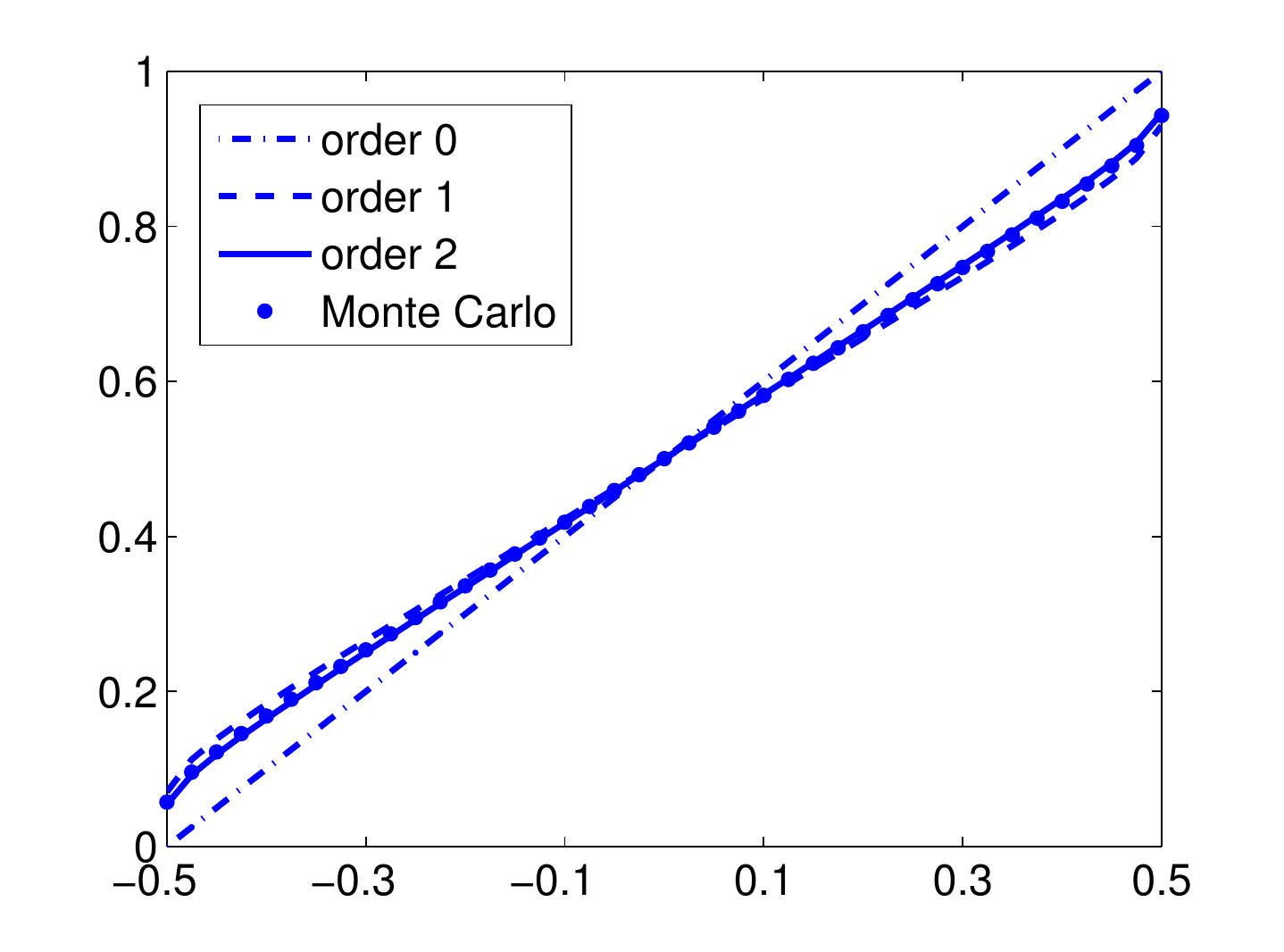}
	\caption{Order 0 (dot-dashed line), order 1 (dashed line) and order 2 (plain line) solutions compared to the solution computed by highly resolved Monte Carlo simulation at $\langle \text{Kn} \rangle=0.1$.}
	\begin{tikzpicture}[overlay]
        		\draw[](1,2.4) node[left]{\rotatebox{0}{ \large $x_1$ }};
        		\draw[](-4.3,6.1) node[left]{\rotatebox{90}{\small  $\displaystyle$ $\frac{T-T_L}{T_R-T_L}$}};
	        
	\end{tikzpicture}
	\label{fig:order1}
\end{figure}
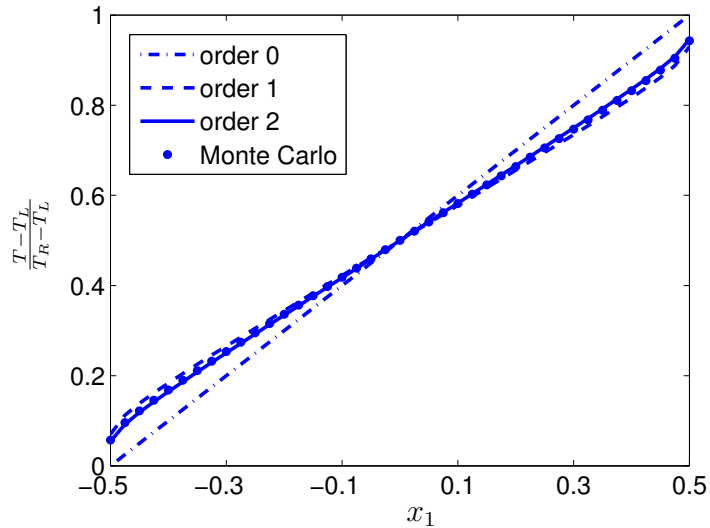

We recall that under the asymptotic analysis, the temperature field is given by
\begin{equation}
T(x_1)=T_{0}(x_1)+\langle \text{Kn} \rangle (T_{G1}(x_1)+T_{K1}(x_1))+ O(\langle \text{Kn} \rangle^2)
\label{secordersol}
\end{equation}
The order 0 solution straightforwardly reads
\begin{equation}
T_{0}(x_1) = \frac{T_L+T_R}{2} + (T_R-T_L)x_1
\end{equation}
since it is the solution of the heat conduction equation subject to no-jump boundary conditions.
Therefore, the boundary conditions for the order 1 field are
\begin{equation}
T_{G1}(x_1=\pm 1/2) = \mp c_1 \frac{\partial T_{0}}{\partial x_1} = \pm c_1 (T_L-T_R)
\end{equation}
which results in
\begin{equation}
T_{G1}(x_1) = 2c_1 (T_L - T_R) x_1
\end{equation}
The boundary layer $(T_R-T_L)\tau_{K1,1}((x_1+1/2)/\langle \text{Kn} \rangle)$ contributes to the solution near the boundary at $x_1=-1/2$, while the function $(T_L-T_R)\tau_{K1,1}((1/2-x_1)/\langle \text{Kn} \rangle)$ contributes close to the boundary at $x_1=1/2$. The resulting  solution correct to order 1 (eq \eqref{secordersol}) is plotted in figure \ref{fig:order1} for $\langle \text{Kn} \rangle=0.1$ in the single relaxation time model and compared to our benchmark (adjoint Monte Carlo \cite{Peraud2014Adjoint}) result. The agreement is excellent; we note in particular that even though the boundary layer correction is small at this Knudsen number, the temperature jumps are considerable and are accurately captured by the asymptotic solution. In contrast, the zeroth order solution (which neglects the temperature jumps) is clearly inadequate.

If desired, calculation of $T(x_1)$ to second order in $\langle \text{Kn} \rangle$ proceeds by solving the heat conduction equation for $T_{G2}$ subject to the second order boundary conditions. Applying \eqref{BC2} to this problem yields
\begin{equation}
T_{G2}(x_1=\pm 1/2)=\mp c_1\frac{\partial T_{G1}}{\partial x_1}=\pm 2c_1^2(T_R-T_L) \label{1D2}
\end{equation}
with the solution
\begin{equation}
T_{G2}(x_1) = 4c_1^2(T_R-T_L)x_1
\label{1D_example_o2}
\end{equation}
The order 2 solution including kinetic boundary layers is also shown in figure \ref{fig:order1} and clearly exhibits improved accuracy with respect to the order 1 solution. 
In fact, in this particular problem where only first derivatives are non zero, the process by which \eqref{1D_example_o2} was derived can be repeated for all orders 
without knowledge of the higher order jump coefficients, leading to an asymptotic solution that is, in principle, correct to all orders. In other words, for $n \geq 1$: $T_{Gn}(x_1) = (-2)^{n}c_1^n(T_R-T_L)x_1$

Summing all orders (provided $2\langle \text{Kn} \rangle c_1< 1$), we obtain:
\begin{equation}
\frac{T_G(x_1)-T_L}{T_R-T_L} = \frac{1}{2}+\frac{x_1}{1+2 \langle \text{Kn} \rangle c_1}
\label{eq:all_orders}
\end{equation}
The boundary layer corrections of all orders can also be obtained (and summed) using the same process. For example, for the boundary at $x_1=-1/2$, we obtain
\begin{equation}
\frac{T_K(x_1)}{T_R-T_L} = \frac{\langle \text{Kn} \rangle}{1+2\langle \text{Kn} \rangle c_1}\tau_{K1,1}\left(\frac{x_1+1/2}{\langle \text{Kn} \rangle}\right).
\label{eq:all_orders_BL}
\end{equation}
The second boundary layer (at $x_1=1/2$) is obtained in an analogous fashion.
This solution is asymptotically accurate to all orders, meaning that the error converges to 0 faster than any power of $\langle \text{Kn} \rangle$; for a discussion on the error associated with the asymptotic expansion see \cite{Sone2007}.

\begin{figure}[htbp]
\centering
	\includegraphics[width=.68\textwidth]{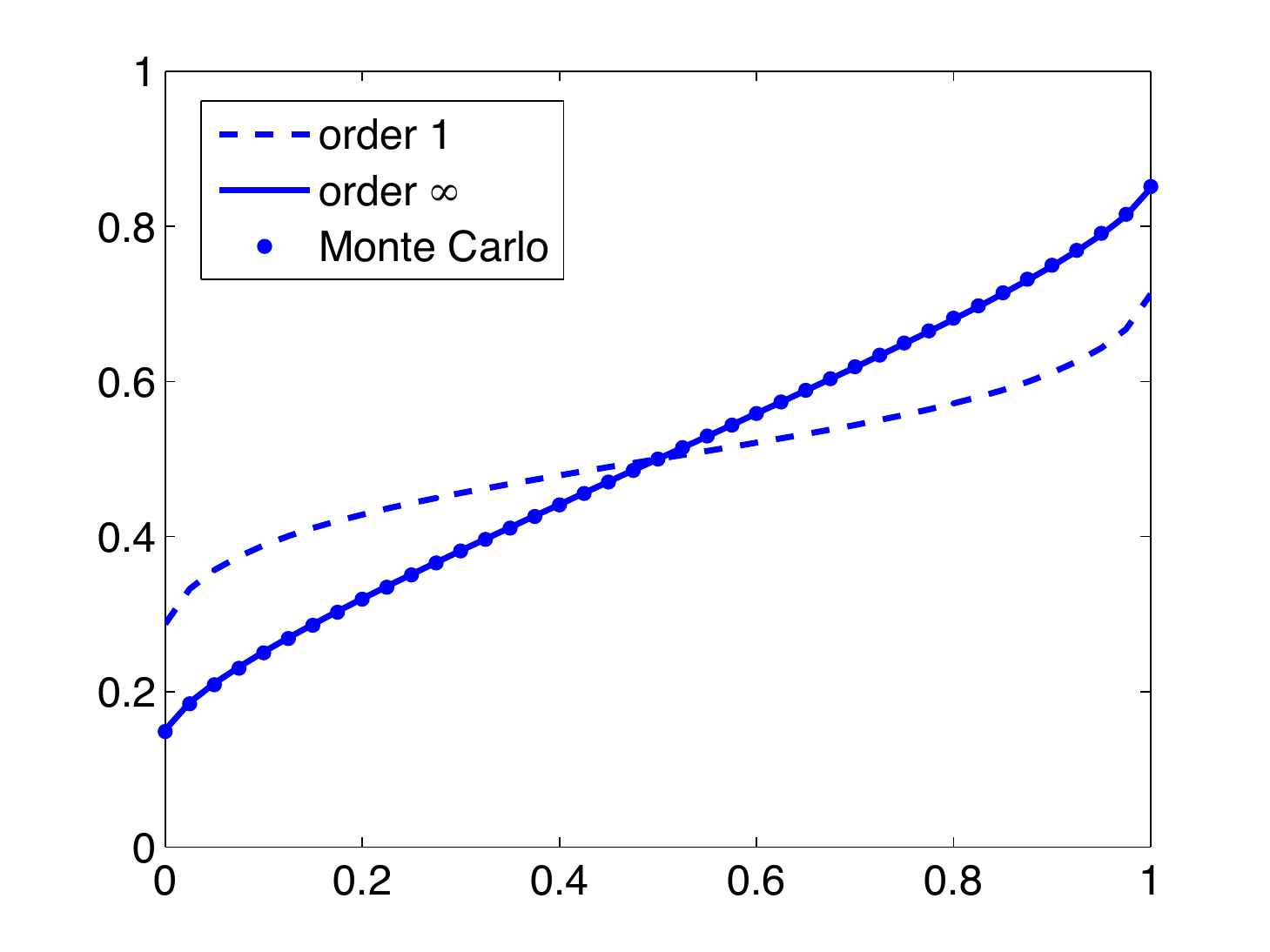}
	\caption{Order 1 solution (dashed line) and ``infinite order'' solution (solid line) compared to the solution computed by a finely resolved Monte Carlo simulation for $\langle \text{Kn} \rangle=0.4$. At this Knudsen number the boundary layer contribution is clearly visible (the solution is no longer a straight line).}
	\begin{tikzpicture}[overlay]
        		\draw[](1,2.7) node[left]{\rotatebox{0}{ \large $x_1$ }};
        		\draw[](-4.3,6.4) node[left]{\rotatebox{90}{\small  $\displaystyle$ $\frac{T-T_L}{T_R-T_L}$}};
	        
	\end{tikzpicture}
	\label{fig:order_inf}
\end{figure}
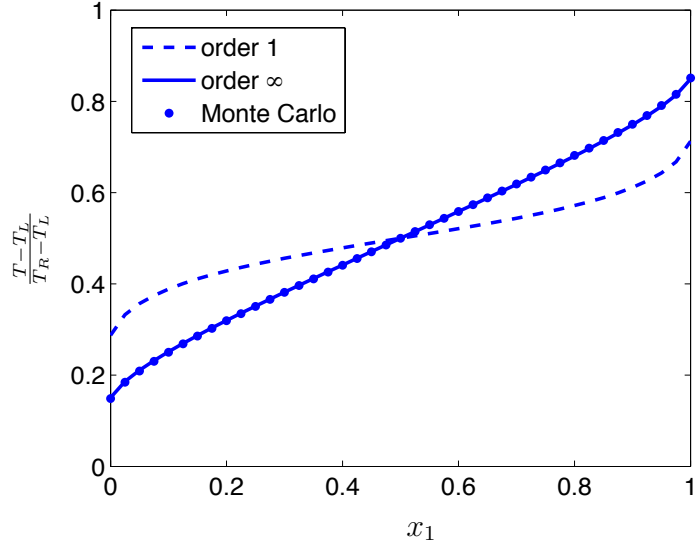

Figure \ref{fig:order_inf}, compares the order 1, infinite order and ``exact'' (Monte Carlo) solution for $\langle \text{Kn} \rangle=0.4$. The infinite order solution is in very good agreement with the exact solution, while the order 1 solution is clearly inadequate at this Knudsen number.

\subsection{``Implicit'' boundary conditions}
\label{implicit}

In the rarefied gas dynamics literature \cite{pof2006} jump boundary conditions are frequently imposed in an ``implicit'' fashion (in the sense that the unknown is on both sides of the equation, 
resulting to what is referred to in the mathematical literature as mixed boundary conditions) thus avoiding the ``stagerred'' solution procedure shown above where the governing equation needs to be solved for each order. For example, a set of boundary conditions up to second order given by 
\begin{equation}
T_{0}\vert_{\mathbf{x}_b}=T_b
\label{alpha}
\end{equation}
\begin{equation}
\left. T_{G1}\right|_{\mathbf{x}_b}=\alpha \left. \frac{\partial T_{0}}{\partial n}\right|_{\mathbf{x}_b}
\label{beta}
\end{equation}
and 
\begin{equation}
\left. T_{G2}\right|_{\mathbf{x}_b}=\alpha \left. \frac{\partial T_{G1}}{\partial n}\right|_{\mathbf{x}_b}+\beta \left. \frac{\partial^2 T_{0}}{\partial n^2}\right|_{\mathbf{x}_b}
\label{gamma}
\end{equation}
may be imposed by solving $\nabla_\mathbf{x}^2 T_G=0$ subject to
\begin{equation}
T_{G} \vert_{\mathbf{x}_b}-T_b = \alpha \langle \text{Kn} \rangle \left. \frac{\partial T_{G}}{\partial n}\right|_{\mathbf{x}_b}+ \beta \langle \text{Kn} \rangle^2 \left. \frac{\partial^2 T_{G}}{\partial n^2}\right|_{\mathbf{x}_b}
\label{eq:implicit}
\end{equation} 
One can show that these two approaches are equivalent (to order $\langle \text{Kn}\rangle^2$) by expanding 
\begin{equation}
T_G \vert_{\mathbf{x}_b}=(T_{0}+\langle \text{Kn} \rangle T_{G1}+\langle \text{Kn} \rangle^2T_{G2}+...)\vert_{\mathbf{x}_b}
\end{equation}
and similarly for $\partial T_G/\partial n|_{\mathbf{x}_b}$ and substituting into (\ref{eq:implicit}). Equating terms of the same orders of $\langle \text{Kn} \rangle$ we obtain equations \eqref{alpha}, \eqref{beta} and \eqref{gamma}, at order zero, one and two, respectively. 

Clearly the implicit form relies on the jump coefficients ($\alpha$, $\beta$, etc) remaining the same at each order (e.g. in \eqref{beta} and \eqref{gamma}).
If the above condition is satisfied, in addition to requiring less solutions of the governing equation, the implicit form has one more advantage: provided higher order derivatives (not included in \eqref{eq:implicit}) do not appear at higher order, the solution will be correct to all orders, since it is easy to verify that \eqref{eq:implicit} then implies that 
\begin{equation}
\left. T_{Gn+2}\right|_{\mathbf{x}_b}=\alpha \left. \frac{\partial T_{Gn+1}}{\partial n}\right|_{\mathbf{x}_b}+\beta \left. \frac{\partial^2 T_{Gn}}{\partial n^2}\right|_{\mathbf{x}_b}
\label{gamma-general}
\end{equation}
for all $n>0$.

This property can be illustrated with the example of section \ref{couette-example}, where $\alpha=c_1$ and $\beta=0$: solution \eqref{eq:all_orders} can be obtained directly by solving $d^2T_G/dx_1^2=0$ subject to 
\begin{equation}
T_{G} \vert_{\mathbf{x}_b}-T_b = c_1 \langle \text{Kn} \rangle \left. \frac{\partial T_{G}}{\partial n}\right|_{\mathbf{x}_b}
\label{eq:implicit-final}
\end{equation}

Although an infinite order solution is always welcome, we also need to keep in mind that some fortuity was involved in this problem in which all higher derivatives of the solution are zero. In the general case, given that $\beta=0$, we expect the implicit condition \eqref{eq:implicit-final} to provide solutions that are accurate at least to second order and at most up to order $m-1$ where $m$ denotes the order of derivative featuring a non-zero jump coefficient. We close by noting that the implicit approach sometimes results in boundary conditions which feature derivatives of the same order as the governing equation which may raise questions about the well-posedness of the mathematical problem. As a resolution to this paradox, we recall that the derivation process followed here (sections \ref{Order1BL} and \ref{Order2BL}) produces the staggered forms of the general type \eqref{alpha}-\eqref{gamma}, which do not present posedness problems. In other words, the implicit form is used merely for convenience and should be discarded if any mathematical/numerical issues arise.

\subsection{A two-dimensional example}
\label{two_dimensional_example}
\begin{figure}[htbp]
\centering
	\includegraphics[width=.7\textwidth]{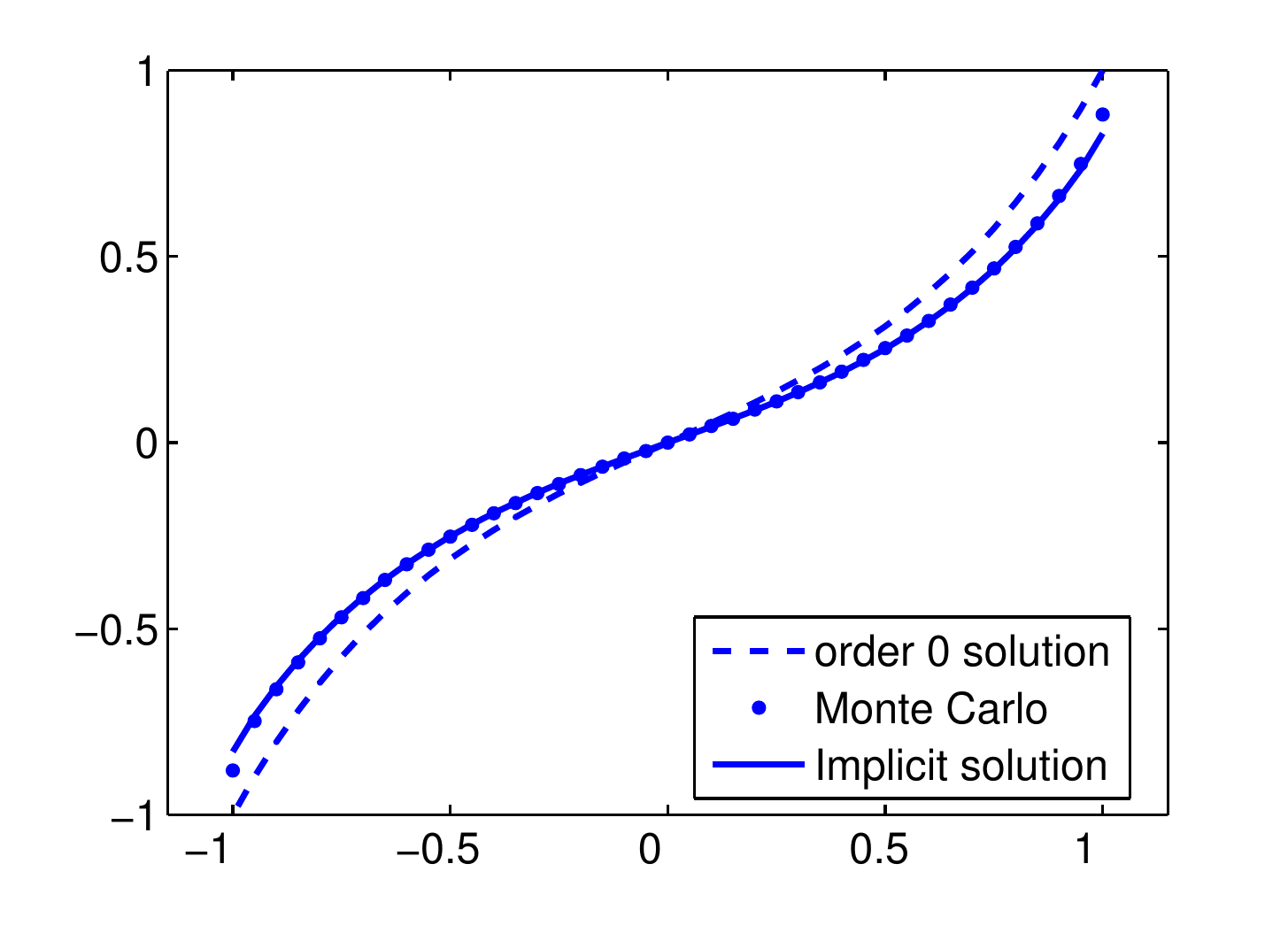}
	\caption{Zeroth order, Monte Carlo and implicit asymptotic  solution for the temperature along the line $x_1=0$ in the two-dimensional example considered in section \ref{two_dimensional_example}, for $\langle \text{Kn} \rangle=0.1$. The inset shows a contour plot of the order 0 solution.}
		\begin{tikzpicture}[overlay]       
        		\draw[](1,2.5) node[left]{\rotatebox{0}{ \large $ x_2 $ }};
                
        		\draw[](-4.3,6.2) node[left]{\rotatebox{90}{\large  $T$ $(K)$}};
	                \node[inner sep=0pt] (2D) at (-1.7,7.7)
                        {\includegraphics[width=.25\textwidth]{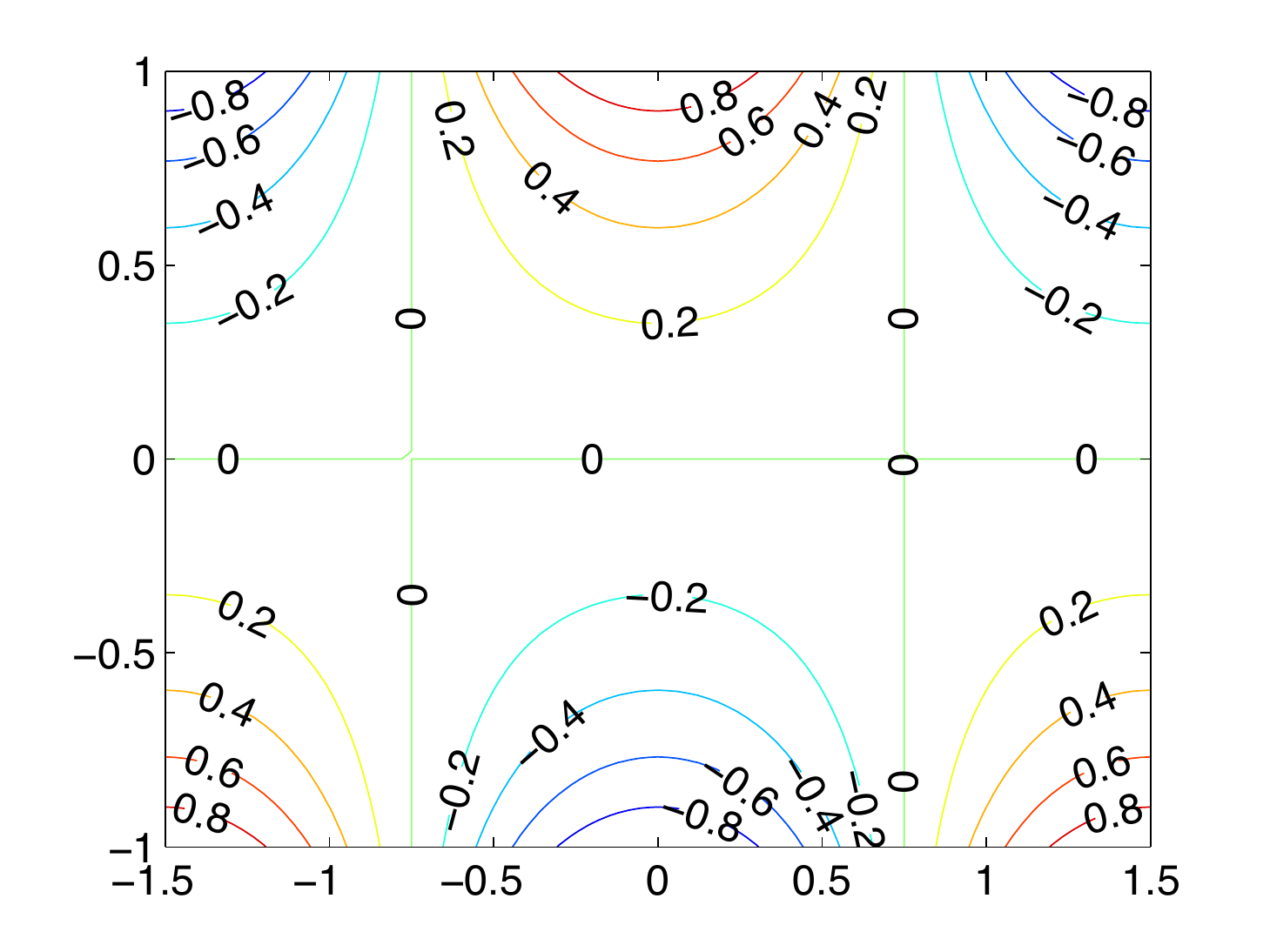}};
                \draw[](-3,7.7) node[left]{\rotatebox{0}{ \tiny $ x_2 $ }};
                \draw[](-1.3,6.3) node[left]{\rotatebox{0}{ \tiny $ x_1 $ }};
                \draw [dashed][red] (-1.64,6.64) -- (-1.64,8.84);
	\end{tikzpicture}
	\label{Tcos_profile}
\end{figure}
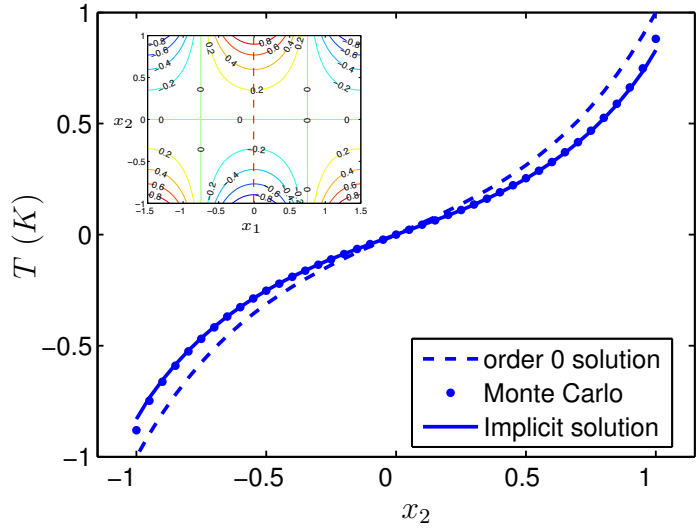

In this section we use a two-dimensional example to illustrate the application as well as convergence properties of the asymptotic solution theory. Specifically, we consider a slab of material that is infinite but subject to a periodic temperature variation in direction $x_1$; the slab has thickness $2L$ in the transverse direction, with the associated dimensionless coordinate $x_2$ defined such that $x_2=0$ describes the median plane of the slab. The material boundaries at $x_2=1$ and $x_2=-1$ are at the prescribed (deviational) temperatures $T_\text{w}\cos(2\pi x_1/3)$ and $-T_\text{w}\cos(2\pi x_1/3)$, respectively. The inset of Figure \ref{Tcos_profile} shows a contour plot of the order 0 solution. 

In what follows, we construct the asymptotic solution of this problem up to $O(\langle \text{Kn}\rangle^2)$, both using the ``order-by-order'' approach and the implicit approach discussed in the previous section. We will then compare these solutions with MC simulation results, both visually along the line $x_1=0$ but also very precisely at location  ($x_1=0,x_2=1$) to compare the order of convergence of the asymptotic solution with the theoretically expected one.

The order 0 solution for the temperature field is given by
\begin{equation}
T_{0}(x_1,x_2) = T_\text{w}\cos \left( \frac{2 \pi x_1}{3} \right) \frac{\sinh \left( \frac{2 \pi x_2}{3} \right)}{\sinh \left( \frac{2 \pi }{3} \right) }
\end{equation}
The order 1 bulk temperature field can be obtained by solving 
the Laplace equation with the boundary conditions:
\begin{equation}
T_{G1}(x_1,x_2 = \pm 1) = \mp c_1 \langle \text{Kn} \rangle \frac{\partial T_{0}}{\partial x_2} (x_1,x_2=\pm 1)
\end{equation}
resulting in
\begin{equation}
\langle \text{Kn} \rangle T_{G1}(x_1,x_2) = -T_\text{w}\langle \text{Kn} \rangle c_1 \frac{2 \pi}{3}   \coth \left( \frac{2 \pi }{3} \right) \cos \left( \frac{2 \pi x_1}{3} \right) \frac{\sinh \left( \frac{2 \pi x_2}{3} \right)}{\sinh \left( \frac{2 \pi }{3} \right) }
\end{equation}
The order 2 bulk temperature field is then obtained by solving the Laplace equation with the boundary conditions:
\begin{equation}
T_{G2}(x_1,x_2 = \pm 1) = \mp c_1 \langle \text{Kn} \rangle \frac{\partial T_{G1}}{\partial x_2}(x_1,x_2=\pm 1) \\
\end{equation}
leading to
\begin{equation}
\langle \text{Kn} \rangle^2 T_{G2}(x_1,x_2) = T_\text{w}\langle \text{Kn} \rangle^2 c_1^2 \left( \frac{2 \pi}{3}   \coth \left( \frac{2 \pi }{3} \right) \right)^2 \cos \left( \frac{2 \pi x_1}{3} \right) \frac{\sinh \left( \frac{2 \pi x_2}{3} \right)}{\sinh \left( \frac{2 \pi }{3} \right) }
\label{second_order_example_pb}
\end{equation}
The solution is complete to second order once the boundary layer contributions are added. The order 1 and order 2 boundary layer correction terms in the vicinity of boundaries $x_2=\pm 1$ are respectively given by:
\begin{equation}
\left \{
\begin{split}
&\langle \text{Kn} \rangle T_{K1}(x_1,x_2) = \mp T_\text{w} \tau_{K1,1}((1 \mp x_2)/\langle \text{Kn}\rangle) \langle \text{Kn} \rangle \frac{2 \pi}{3}  \coth \left( \frac{2 \pi }{3} \right)   \\
&\langle \text{Kn} \rangle^2 T_{K2}(x_1,x_2) = \pm T_\text{w} \tau_{K1,1}((1 \mp x_2)/\langle \text{Kn}\rangle) c_1 \langle \text{Kn} \rangle^2 \left( \frac{2 \pi}{3}   \coth \left( \frac{2 \pi }{3} \right) \right)^2
\end{split}
\right.
\end{equation}

As explained in the previous section, a solution of a similar order can be achieved by directly looking for the solution of the Laplace equation $T_G$  with boundary conditions:
\begin{equation}
T_{G}(x_1,x_2 = \pm 1) = \mp c_1 \langle \text{Kn} \rangle \frac{\partial T_{G}}{\partial x_2}(x_1,x_2=\pm 1) 
\end{equation}
This is the case here because, as shown in section \ref{Order2BL_prescribed}, second-order derivatives do not appear in the jump conditions or the temperature boundary layer.
Applying these ``implicit'' boundary conditions, we obtain
\begin{equation}
T_{G}(x_1,x_2) = \frac{T_\text{w}}{1 + c_1\langle \text{Kn} \rangle \frac{2 \pi}{3} \coth \left(\frac{2 \pi}{3}\right) } \cos \left( \frac{2 \pi x_1}{3}\right) \frac{\sinh \left( \frac{2 \pi x_2 }{3} \right) }{\sinh \left( \frac{2 \pi }{3} \right) }
\end{equation}
The kinetic boundary layer corrections in the vicinity of the boundaries at $x_2=\pm 1$ are given by $\mp \tau_{K1,1}((1 \mp x_2)/\langle \text{Kn}\rangle)\langle \text{Kn} \rangle \partial T_G/\partial x_2 (x_1=0,x_2=\pm 1) $. Evaluating the combined (bulk and boundary layer correction) solution at $(x_1=0,x_2=1)$, we obtain 
\begin{equation}
T_\text{w} \frac{1-\tau_{K1,1}(0) \langle \text{Kn} \rangle \frac{2 \pi}{3} \coth \left( \frac{2 \pi }{3} \right)}{1 + c_1 \langle \text{Kn} \rangle \frac{2 \pi}{3}\coth \left( \frac{2 \pi }{3} \right)}
\label{infinite_order}
\end{equation}

This solution is compared to a highly-resolved MC simulation result in Fig. \ref{Tcos_profile} for the case $\langle \text{Kn}\rangle =0.1$. The material model used is the single-relaxation-time model defined in section \ref{BL_results}. The MC solution was obtained using the adjoint Monte Carlo method described in \cite{Peraud2014Adjoint} and will be denoted $T_{MC}$ below. 

\begin{figure}[htbp]
\centering
	\includegraphics[width=.7\textwidth]{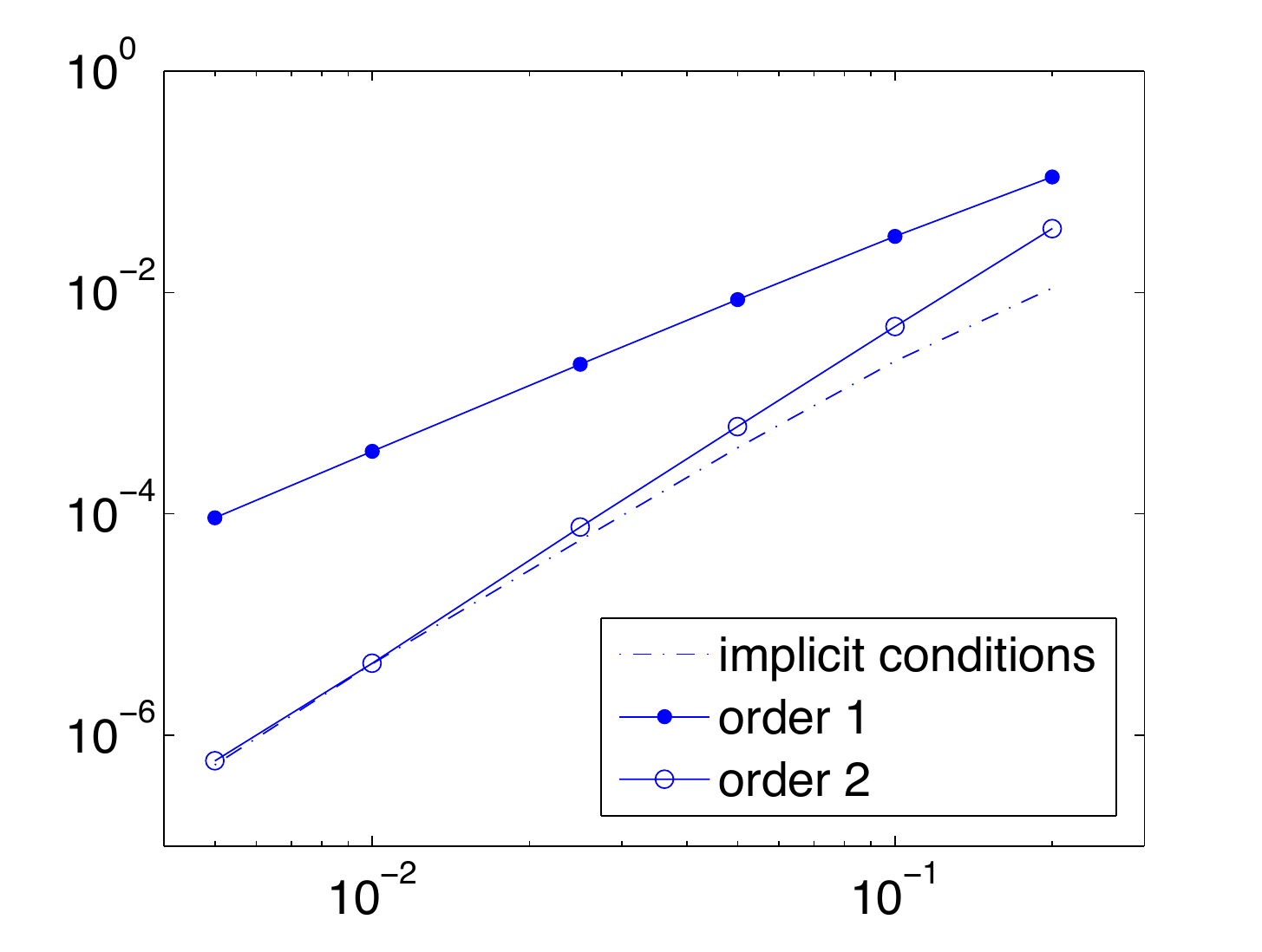}
	\caption{Convergence of asymptotic temperature solutions at  $(x_1=0,x_2=1)$ in the two-dimensional example considered in section \ref{two_dimensional_example}.}
		\begin{tikzpicture}[overlay]
        		\draw[](1,2) node[left]{\rotatebox{0}{ \large $\langle \text{Kn} \rangle $ }};
        		\draw[](-4.6,6) node[left]{\rotatebox{90}{\large  $\displaystyle$ $\epsilon_T$ $(K)$}};
	        
	\end{tikzpicture}
	\label{cos_sinh_o2}
\end{figure}
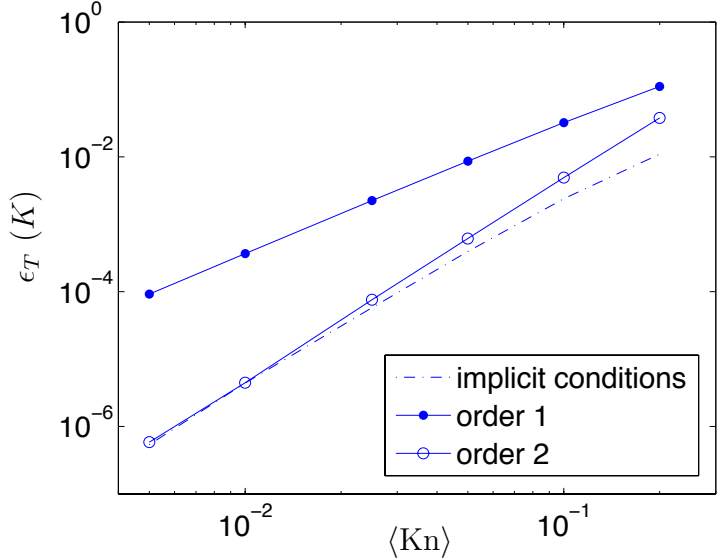

Figure \ref{cos_sinh_o2} plots $|T_{MC}(x_1=0,x_2=1)-T_\text{asymptotic}(x_1=0,x_2=1)|$ for 3 asymptotic solutions, namely, the first-order solution $T_0+\langle \text{Kn}\rangle(T_{G1}+T_{K1})$, the second-order solution $T_0+\langle \text{Kn}\rangle(T_{G1}+T_{K1})+ \langle \text{Kn}\rangle^2 (T_{G2}+T_{K2})$,  and the implicit solution \eqref{infinite_order}. The figure shows that the implicit formulation leads to an order 2 solution overall which additionally features slightly improved accuracy compared to the ``regular'' order 2 solution. As explained in section \ref{implicit}, the solution would be ``infinite'' order if no higher order derivative appeared in the jump boundary conditions. 
The third-order convergence observed for the implicit solution seems to suggest that a non-zero jump coefficient appears in front of the third-order derivative ($m=3$).

\section{Extension to time-dependent problems}
\label{time-dep}
Although the analysis presented here has so far been limited to steady problems, extension to unsteady problems is relatively straightforward. In the field of rarefied gas dynamics the Hilbert expansion has been extended to time-dependent problems by Sone \cite{Sone2002} and Takata \cite{Takata_2012,takata2015}, who showed that, other than the additional time-derivative in the governing equation, time dependence does not introduce any new physics up to order 1 in $\langle \text{Kn}\rangle$.  

In this section we show that this is also true for phonon transport for the case of prescribed temperature boundaries by introducing the dimensionless time-dependent Boltzmann equation
\begin{equation}
\text{St}\frac{\partial \Phi}{\partial t} + \boldsymbol{\Omega} \cdot \nabla_{\mathbf{x}} \Phi = \frac{\mathcal{L}(\Phi) -\Phi}{\text{Kn}}
\end{equation}
where $t$ is a dimensionless time, defined by $t\equiv t'/t_0$, where $t_0$ is a characteristic time of variation and the Strouhal number is given by
\begin{equation}
\text{St}_{\omega,p}=\text{St}=\frac{L}{V_g t_0}
\end{equation}
We analyze cases where $\langle \text{St}\rangle \sim \langle \text{Kn} \rangle$, where the average Strouhal number, $\langle \text{St}\rangle$, follows from an analogous definition to that of $\langle \text{Kn}\rangle$ in \eqref{Kn_definition}. The condition $\langle \text{St}\rangle \sim \langle \text{Kn} \rangle$ can be rewritten as $t_0\sim L^2/\kappa\sim \langle \tau \rangle/\langle \text{Kn} \rangle ^2$, which implies an assumption of diffusive scaling in time.

Expanding the time dependent function $\Phi$ as in Eq. \eqref{phi_expansion} results in the same forms for orders 0 and 1 (equations \eqref{phig0} to \eqref{phig1}). Differences appear at order 2. Specifically, the form of the order 2 solution reads:
\begin{equation}
\Phi_{G2} =\mathcal{L}(\Phi_{G2}) - \frac{\text{Kn}}{\langle \text{Kn} \rangle} \boldsymbol{\Omega} \cdot \nabla_{\mathbf{x}} T_{G1} -\frac{\text{St} \text{Kn}}{\langle \text{Kn} \rangle^2}\frac{\partial T_{0}}{\partial t}  + \frac{\text{Kn}^2}{\langle \text{Kn} \rangle^2}\boldsymbol{\Omega} \cdot \nabla_{\mathbf{x}} \left( \boldsymbol{\Omega} \cdot \nabla_{\mathbf{x}} T_{0} \right)
\label{phi_g2_time}
\end{equation}
Applying the solvability condition \eqref{eq:solvability} results in
\begin{equation}
\int_{\omega,p,\boldsymbol{\Omega}}  \frac{C_{\omega,p}}{4 \pi \tau} 
\left( 
\frac{\text{St} \text{Kn}}{\langle \text{Kn} \rangle^2} \frac{\partial T_0}{\partial t} +\frac{\text{Kn}}{\langle \text{Kn} \rangle} \boldsymbol{\Omega} \cdot \nabla_{\mathbf{x}} T_{G1}
 - \frac{\text{Kn}^2}{\langle \text{Kn} \rangle^2}\boldsymbol{\Omega} \cdot \nabla_{\mathbf{x}} \left( \boldsymbol{\Omega} \cdot \nabla_{\mathbf{x}} T_0 \right)
\right) d^2 \boldsymbol{\Omega} d \omega =0
\end{equation}
which, after integration, yields the heat equation for the order 0 temperature field:
\begin{equation}
\frac{\partial T_{0}}{\partial t'} = \frac{\kappa}{C} \nabla_{\mathbf{x}'}^2 T_{0}.
\label{transheat}
\end{equation}
Applying the solvability condition to the order 3 solution similarly yields the heat equation for the order 1 temperature field. Although not strictly needed for our purpose here, we may solve for $\mathcal{L}(\Phi_{G2})$ in Eq. \eqref{phi_g2_time} by writing:
\begin{equation}
T_{G2} =  \frac{1}{C} \int_{\omega,p,\boldsymbol{\Omega}} \frac{C_{\omega,p}}{4 \pi} \Phi_{G2} d^2 \boldsymbol{\Omega} d \omega
\end{equation}
which, combined with \eqref{phi_g2_time}, yields
\begin{equation}
 \mathcal{L}(\Phi_{G2}) = T_{G2}+\frac{1}{C\langle \text{Kn} \rangle^2} \int_{\omega,p} C_{\omega,p} \text{St} \text{Kn} d \omega\frac{\partial T_{0}}{\partial t}-\frac{1}{C\langle \text{Kn} \rangle^2} \int_{\omega,p} C_{\omega,p} \frac{\text{Kn}^2}{3} d \omega \nabla_\mathbf{x}^2 T_{0}
\end{equation}
which in the general case differs from $T_{G2}$. We note that $\mathcal{L}(\Phi_{G2})=T_{G2}$ holds in the case where the relaxation time does not depend on frequency and polarization.

The order 0 boundary condition was obtained in section \ref{boundary_conditions} by noticing that the order 0 distribution matches the distribution emitted by the boundary with no boundary layer correction. Introducing time dependence does not modify this result. { \it Therefore the Dirichlet boundary condition $T_{G0}= T_b$ remains unmodified at order 0 in the time-dependent case}. At order 1, we showed that the jump boundary condition emerges from the analysis of the boundary layer correction required by the mismatch between the order 1 bulk distribution and the boundary emitted distribution. As before, time-dependence does not modify the form of the order 1 bulk distribution. Therefore,  {\it the order 1 jump condition \eqref{BC1} remains unmodified in the presence of  time dependence}. Similarly, the derivation of the order 0 condition for diffuse reflective walls results from an order 1 analysis. {\it The Neumann condition \eqref{order0_diffuse} is unmodified}. The order 2 boundary layer analysis presented in section \ref{diffuse_order1} that yields condition \eqref{diffuse_O1_condition} requires a modification since the relation $\nabla^2_\mathbf{x} T_0 = 0$ is replaced by the diffusion equation. In this work, we did not proceed to analyze in detail how the order 1 boundary condition for diffuse reflective walls is modified. 


This shows that the theory developed in this article may be applied to time-dependent problems (exhibiting diffusive scaling in time) up to order 1 in the presence of prescribed temperature boundaries, with the only change being that the Laplace equation is replaced by the unsteady heat equation \eqref{transheat}. 

\subsection{Application to a transient problem}
\begin{figure}
\begin{center}
\includegraphics[width=.5\textwidth]{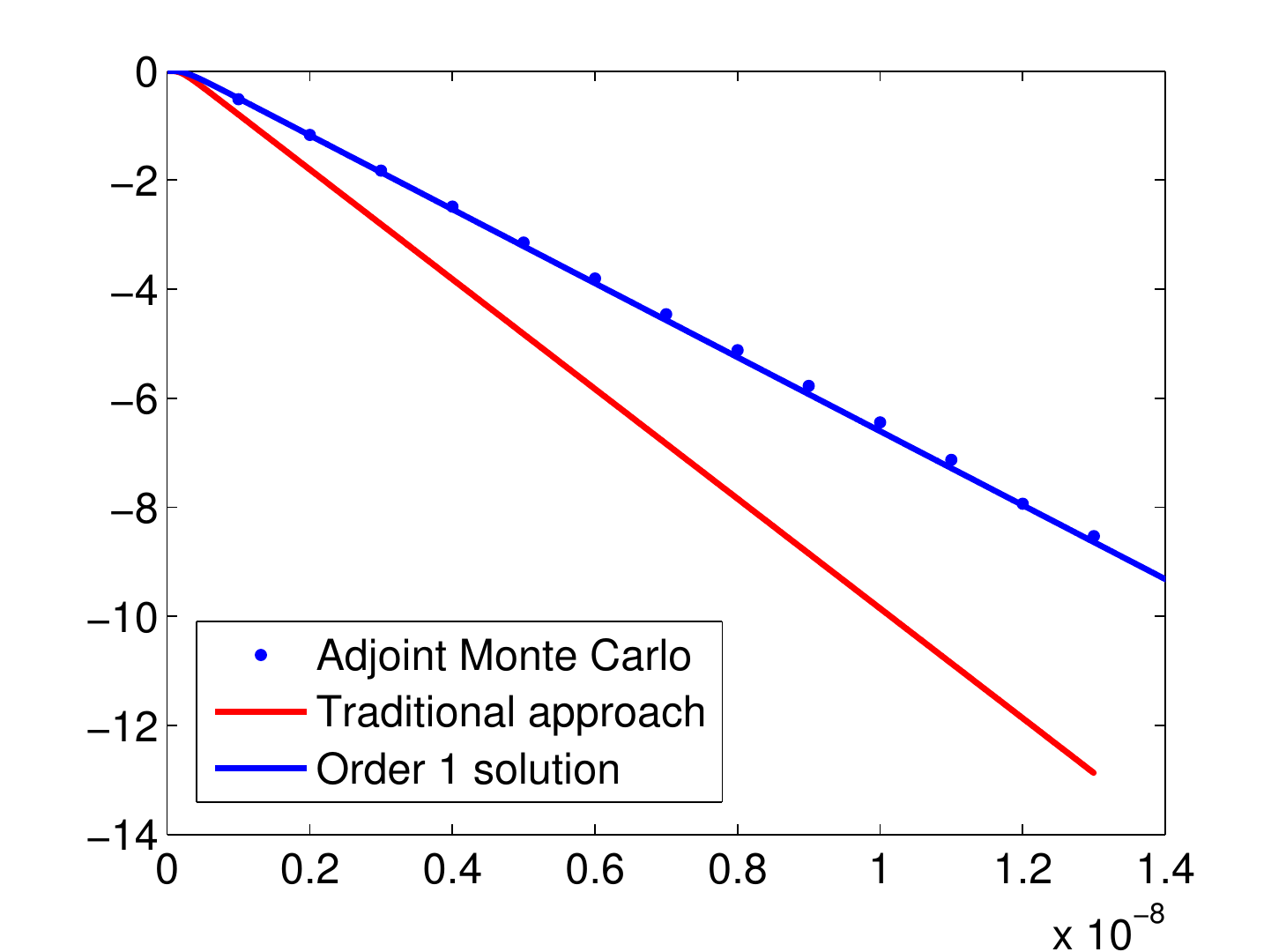}
\caption{Temperature at the center of a square particle after initial heating.}
\begin{tikzpicture}[overlay]
\draw[](.9,1.3) node[left]{\rotatebox{0}{\small t(s)}};
\draw[](-3,4.1) node[left]{\rotatebox{90}{\small $\log(\left|\hat{T}(t)-T_{b}\right|)$}};
\end{tikzpicture}
	\label{Dirichlet_and_slip_01}
\end{center}
\end{figure}
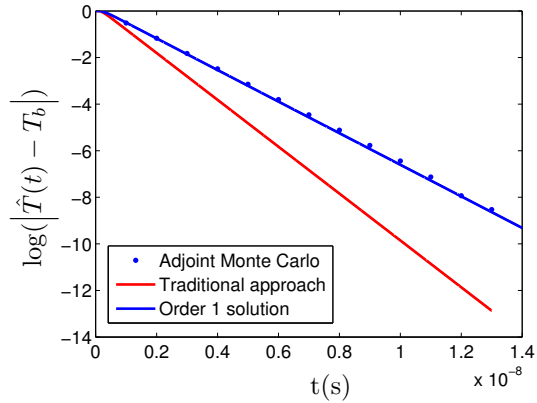

To illustrate and briefly validate some of the conclusions of the previous section, we consider here a square particle heated to a uniform temperature of 301 K and placed in a thermal bath at 300 K, such that its boundary is well described by a prescribed temperature of $T_{b}=300$ K. We also assume that the Knudsen number is small  such that we can calculate the temperature field inside the particle by solving the heat equation
\begin{equation}
\frac{\partial T}{\partial t'} = \frac{\kappa}{C} \nabla_{\mathbf{x}'}^2 T  
\end{equation}
with the first-order boundary conditions derived in this work. For convenience we use the "implicit form" described in section \ref{implicit}
\begin{equation}
T(\mathbf{x}=\mathbf{x}_\text{b})-T_{b} = c_1 \langle \text{Kn} \rangle \frac{\partial T}{\partial n}
\end{equation}
In Figure \ref{Dirichlet_and_slip_01}, we show a measure of this temperature relaxation process, namely $\left|\hat{T}(t)-T_{b}\right|$, for $\langle \text{Kn} \rangle =0.1$, where $\hat{T}(t)$ denotes the temperature at the center of the particle. The heat equation solution was obtained using a finite difference scheme. Here we note that the particle center is sufficiently far from the boundary that no kinetic boundary layer correction is required. The material model adopted here is that of silicon with a single relaxation time ($c_1\approx1.13$).

This solution is compared with results obtained using the adjoint Monte Carlo method presented in Ref. \cite{Peraud2014Adjoint}. We also show the solution obtained from the (traditional) heat equation with the Dirichlet boundary conditions $T(\mathbf{x}=\mathbf{x}_\text{b})=300$K.  
The figure shows that the asymptotic solution is in excellent agreement with the MC solution, while, as expected, the traditional approach (with Dirichlet boundary conditions)--which corresponds to the zeroth-order solution--significantly overpredicts the particle cooling rate. 

\section{Application to interfaces between materials}
\label{kapitza}

The theoretical and numerical considerations presented in this paper are quite general and can be extended to a variety of problems where boundaries introduce ``size effects'' by injecting inhomogeneity into the problem. A classic example of such a problem is the interface between two materials: the presence of the interface results in a temperature jump, already shown in this work to be the signature of the kinetic correction required due to the inhomogeneity associated with the presence of a boundary.
In this section we show how the asymptotic theory enables us to  rigorously relate the Kapitza conductance to the kinetic properties of the interface (e.g. reflection/transmission coefficients).
Our aim here is not to conduct an exhaustive study but rather to demonstrate the applicability of the ideas presented earlier. As a result, we will focus on one specific transmission model and the single relaxation time model. We assume the following:
\begin{itemize}
\item[-] The interface separating the two media, denoted $a$ and $b$, is sharp (infinitely thin) and planar. 
\item[-] When a phonon encounters the interface, it is either reflected or transmitted. In either case, its traveling direction is randomized while it keeps the same frequency and polarization. We denote the transmission probability from material $a$ to material $b$ by  $\chi_{ab}$, while $\rho_{ab}=1-\chi_{ab}$ denotes the probability of reflection at the interface while traveling from $a$ to $b$. Similarly, $\chi_{ba}$ and $\rho_{ba}$ denote the transmission and reflection probabilities for travel from $b$ to $a$, respectively. 
\end{itemize}
In what follows, we will use $\tau_a$, $C_{\omega,p,a}$, $V_{g,a}$ and $\tau_b$, $C_{\omega,p,b}$, $V_{g,b}$ to denote the relaxation time, frequency-dependent specific heat and magnitude of the (frequency and polarization dependent) group velocity in materials $a$ and $b$, respectively.
As before and without loss of generality, let us align the interface with the $x_2-x_3$ plane (at $x_1=0$) and let the positive $x_1$ direction point from material $a$ to material $b$. In this notation, the kinetic boundary condition associated with the interface is given by 
\begin{equation} \label{interface_condition}
\left \{
\begin{split}
V_{g,b} \frac{C_{\omega,p,b}}{4} \Phi_b^+\vert_{x_1=0^+}= \int_{\Omega_1>0} \chi_{ab} \Phi_a^+ \vert_{x_1=0^-}\frac{C_{\omega,p,a}}{4\pi} & V_{g,a} \Omega_1 d^2 \boldsymbol{\Omega} \\
 &- \int_{\Omega_1<0} \rho_{ba} \Phi_b^- \vert_{x_1=0^+}\frac{C_{\omega,p,b}}{4\pi} V_{g,b} \Omega_1 d^2 \boldsymbol{\Omega} \\
V_{g,a} \frac{C_{\omega,p,a}}{4} \Phi_a^-\vert_{x_1=0^-}=-\int_{\Omega_1<0} \chi_{ba} \Phi_b^- \vert_{x_1=0^+}\frac{C_{\omega,p,b}}{4 \pi} & V_{g,b} \Omega_1 d^2 \boldsymbol{\Omega} \\
 &+ \int_{\Omega_1>0} \rho_{ab} \Phi_a^+ \vert_{x_1=0^-}\frac{C_{\omega,p,a}}{4 \pi} V_{g,a} \Omega_1 d^2 \boldsymbol{\Omega}
\end{split}
\right.
\end{equation}
where superscript ``$^+$'' (resp.  ``$^-$'') refers to particles moving in the positive (resp. negative) $x_1$ direction. 

The order 0 solution in each material phase is solution to the Laplace equation $\nabla^2_\mathbf{x} T_{0}=0$ with the condition $T_{G0,a}\vert_{x_1=0^-}=T_{G0,b}\vert_{x_1=0^+}=T_{0}\vert_{x_1=0}$ at the interface. Replacing $\Phi_a\vert_{x_1=0^-}$ and $\Phi_b\vert_{x_1=0^+}$ by $T_{0}\vert_{x_1=0}$ in \eqref{interface_condition} and performing the integrations, we obtain:
\begin{equation} \label{interface_condition_0_1}
\left \{
\begin{split}
&V_{g,b} C_{\omega,p,b} T_{0}\vert_{x_1=0}=\chi_{ab} T_{0}\vert_{x_1=0} C_{\omega,p,a} V_{g,a} + \rho_{ba} T_{0} \vert_{x_1=0} C_{\omega,p,b} V_{g,b}\\
&V_{g,a} C_{\omega,p,a} T_{0}\vert_{x_1=0}=\chi_{ba} T_{0}\vert_{x_1=0} C_{\omega,p,b} V_{g,b} + \rho_{ab} T_{0}\vert_{x_1=0} C_{\omega,p,a} V_{g,a} 
\end{split}
\right.
\end{equation}
which implies
\begin{equation}
0=\chi_{ab} C_{\omega,p,a} V_{g,a} - \chi_{ba} C_{\omega,p,b} V_{g,b}.
\label{detailed_balance}
\end{equation}
The principle of detailed balance guarantees that the above is true for all $\omega,p$. Note that the condition $T_{G0,a}\vert_{x_1=0^-}=T_{G0,b}\vert_{x_1=0^+}=T_{0}\vert_{x_1=0}$ does not determine the value of $\Phi_{0}\vert_{x_1=0}=T_{0}\vert_{x_1=0}$. The additional required condition is given by heat flux continuity:
\begin{equation}
\kappa_a \left. \frac{\partial T_{0}}{\partial x_1}\right|_{x_1=0^-} = \kappa_b \left. \frac{\partial T_{0}}{\partial x_1}\right|_{x_1=0^+}
\end{equation}

Following the procedure of section \ref{boundary_conditions}, we find that the order 1 solutions
\begin{equation} 
\left \{
\begin{split}
\Phi_{1,a}=T_{1,a} -\frac{\text{Kn}_a}{\langle \text{Kn} \rangle} \boldsymbol{\Omega} \cdot \nabla T_{0,a}\\
\Phi_{1,b}=T_{1,b} -\frac{\text{Kn}_b}{\langle \text{Kn} \rangle} \boldsymbol{\Omega} \cdot \nabla T_{0,b}
\end{split}
\right.
\end{equation}
cannot satisfy condition \eqref{interface_condition} without the introduction of boundary layers. Here $\text{Kn}_i$ denotes $V_{g,i}\tau_i/L$, while $\langle \text{Kn}\rangle$ is a ``reference'' Knudsen number calculated from the properties of one of the two materials (results are independent of the chosen reference).

 We introduce two boundary layer functions $\Psi_{Ka}$ and $\Psi_{Kb}$, and two constants $c_a$ and $c_b$, anticipating temperature jumps from the order 0 at the interface of the form 
\begin{equation} 
\left\{
\begin{split}
\left. T_{1,a}\right|_{x_1=0^-}=c_a \left. \frac{\partial T_{0,a}}{\partial n_a}\right|_{x_1=0^-}\\
\left. T_{1,b}\right|_{x_1=0^+}=c_b \left. \frac{\partial T_{0,b}}{\partial n_b}\right|_{x_1=0^+}
\end{split}
\right.
\end{equation}
Limiting our analysis to variations only in the $x_1$ direction, we insert the order 1 solution (boundary layer included) in condition \eqref{interface_condition}, to obtain
\begin{equation} \label{interface_condition2}
\left \{
\begin{split}
&V_{g,b} \frac{C_{\omega,p,b}}{4}  \left( -\text{Kn}_b \Omega_1+c_b \langle \text{Kn} \rangle +\Psi_{Kb} \langle \text{Kn} \rangle \right)\vert_{x_1=0^+} \left. \frac{\partial T_{0,b}}{\partial x_1}\right|_{x_1=0^+} \\
 & =\int_0^{1}\chi_{ab}  \frac{C_{\omega,p,a}}{2} V_{g,a}\left( -\text{Kn}_a \Omega_1' -c_a \langle \text{Kn} \rangle -\Psi_{Ka} \langle \text{Kn} \rangle \right)\vert_{x_1=0^-} \left. \frac{\partial T_{0,a}}{\partial x_1} \right|_{x_1=0^-} \Omega_1' d \Omega_1' \\ 
 &-  \int_{-1}^{0} \rho_{ba}  \frac{C_{\omega,p,b}}{2} V_{g,b}  \left( -\text{Kn}_b \Omega_1' +c_b \langle \text{Kn} \rangle +\Psi_{Kb} \langle \text{Kn} \rangle \right)\vert_{x_1=0^+} \left.\frac{\partial T_{0,b}}{\partial x_1}\right|_{x_1=0^+} \Omega_1' d \Omega_1'   \\
&V_{g,a} \frac{C_{\omega,p,a}}{4} \left( -\text{Kn}_a \Omega_1 -c_a \langle \text{Kn} \rangle -\Psi_{Ka} \langle \text{Kn} \rangle \right)\vert_{x_1=0^-}\left.\frac{\partial T_{0,a}}{\partial x_1} \right|_{x_1=0^-}\\
&=-\int_{-1}^{0} \chi_{ba} \frac{C_{\omega,p,b}}{2} V_{g,b} \left( -\text{Kn}_b \Omega_1' +c_b \langle \text{Kn} \rangle +\Psi_{Kb} \langle \text{Kn} \rangle \right)\vert_{x_1=0^+} \left. \frac{\partial T_{0,b}}{\partial x_1}\right|_{x_1=0^+} \Omega_1' d \Omega_1' \\
&  +\int_{0}^{1} \rho_{ab}  \frac{C_{\omega,p,a}}{2} V_{g,a} \left( -\text{Kn}_a \Omega_1' -c_a \langle \text{Kn} \rangle -\Psi_{Ka} \langle \text{Kn} \rangle \right)\vert_{x_1=0^-} \left. \frac{\partial T_{0,a}}{\partial x_1}\right|_{x_1=0^-}  \Omega_1' d \Omega_1'
\end{split}
\right.
\end{equation}
We then solve this boundary layer problem numerically to obtain the condition 
\begin{equation}
T_{1,b}-T_{1,a} =  \tilde{c} \kappa_a \left. \frac{\partial T_{0,a}}{\partial x}\right|_{x_1=0^-} \langle \text{Kn} \rangle
\end{equation}
with $\tilde{c} = c_a/\kappa_a+c_b/\kappa_b$,  
describing the first-order temperature jump across the interface.
The numerical procedure used is described in \cite{JPThesis}.

\subsection{Validation}
We test the asymptotic solution method outlined here on a simple one-dimensional problem with the following features:
\begin{itemize}
\item[-] The total length of the system is $2L$. The two materials are aluminum ($-1\leq x_1<0$, hence, material $a$) and silicon ($0<x_1\leq 1$, hence, material $b$). Here, we emphasize that we perform this calculation to {\it validate} asymptotic theory describing {\it phonon transport} across the interface. As a result, the aluminum model used here does not include electronic transport, which leads to $\kappa_\text{Al}=27.7$ W/mK. The choice of aluminum was motivated by the fact that this metal is frequently used as a transducer in transient thermoreflectance experiments \cite{Minnich2011, zeng2014, Hu_2015, Zeng_2015} and thus a reliable and well understood Monte Carlo simulation model -- a priority for validation studies -- exists \cite{prb} for this material. We use $\mathcal{I}_{\text{Al}}$ and $\mathcal{I}_{\text{Si}}$ to denote the range of frequencies of the two material dispersion relations, respectively. We also use a constant relaxation time model in each material; specifically, we take $\tau_a = 10^{-11}$s in Al and $\tau_b = 4 \times 10^{-11}$ s in Si.
\item[-] A temperature difference of 1 K is applied across the system by imposing a prescribed temperature of 301 K at $x_1=-1$, while the boundary at $x_1=1$ is maintained at 300 K. We note that the prescribed temperatures are used here to impose a temperature gradient onto the system. They are in no way linked to the interface model.
\item[-] We define $\langle \text{Kn} \rangle$ as the ratio between the mean free path in the silicon phase and $L$. We choose $L$ such that $\langle \text{Kn} \rangle=0.1$. 
\item[-] The phonon transmissivities at the interface $x_1=0$ are adapted from the model described in \cite{minnichthesis}, which given a ``target'' interface conductance $G$ (as input), predicts  
\begin{equation}
\chi_{ab} = \frac{\frac{2}{\int_{\omega \in \mathcal{I}_{\text{Al}} \cap \mathcal{I}_{\text{Si}},p} C_{\omega,p,\text{Al}} V_{g,\text{Al}}}}{\frac{1}{\int_{\omega \in \mathcal{I}_{\text{Al}},p} C_{\omega,p,\text{Al}} V_{g,\text{Al}}}+\frac{1}{\int_{\omega \in \mathcal{I}_{\text{Si}},p} C_{\omega,p,\text{Si}} V_{g,\text{Si}}}+\frac{1}{2G}}
\end{equation}
for frequencies in $\mathcal{I}_{\text{Al}} \cap \mathcal{I}_{\text{Si}}$ (0 otherwise). Coefficients $\chi_{ba}$ are deduced from the principle of detailed balance.
\end{itemize}

Due to the one-dimensional nature of the problem studied here and the absence of higher than first-order derivatives of temperature in either material, an ``infinite'' order solution is possible: it can be obtained by solving the following system of four equations in four unknowns ($T_\text{Al}(x_1=-1)$, $T_\text{Al}(x_1=0^-)$, $T_\text{Si}(x_1=0^+)$ and $T_\text{Si}(x_1=1)$)
\begin{equation}
\left \{
\begin{split}
&1-T_\text{Al}(x_1=-1)=c_{\text{Al}} \langle \text{Kn} \rangle \left( T_\text{Al}(x_1=0^-)-T_\text{Al}(x_1=-1) \right)\\
&T_\text{Si}(x_1=0^+)-T_\text{Al}(x_1=0^-) = \tilde{c} \langle \text{Kn} \rangle\kappa_\text{Si} \left. \frac{\partial T_\text{Si}}{\partial x_1} \right|_{x_1=0^+} \\
&\kappa_\text{Al} \left. \frac{\partial T_\text{Al}}{\partial x_1} \right|_{x_1=0^-} =\kappa_\text{Si} \left. \frac{\partial T_\text{Si}}{\partial x_1} \right|_{x_1=0^+} \\
&T_\text{Si}(x_1=1)=c_{\text{Si}} \langle \text{Kn} \rangle \left( T_\text{Si}(x_1=0^+)-T_\text{Si}(x_1=1) \right)\label{inter}
\end{split}
\right.
\end{equation}
We emphasize here that the temperature jump relations at $x=\pm 1$ (first and last lines in \eqref{inter}) appear only because of the particular formulation used here for imposing the temperature gradient, namely using prescribed temperature boundaries far from the interface. Here, but also in general, the dynamics of the interface are solely described by the second and third lines of \eqref{inter}, namely heat flux continuity and the temperature jump across the interface.

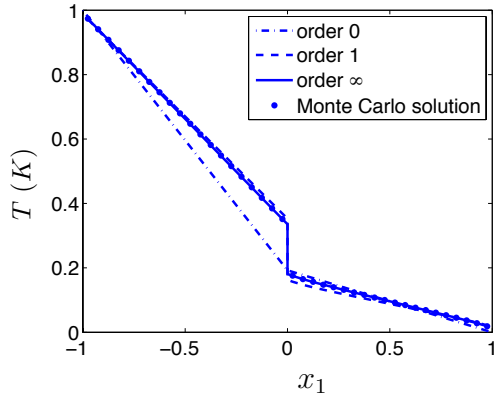
\begin{figure}[htbp]
\centering
	\includegraphics[width=.48\textwidth]{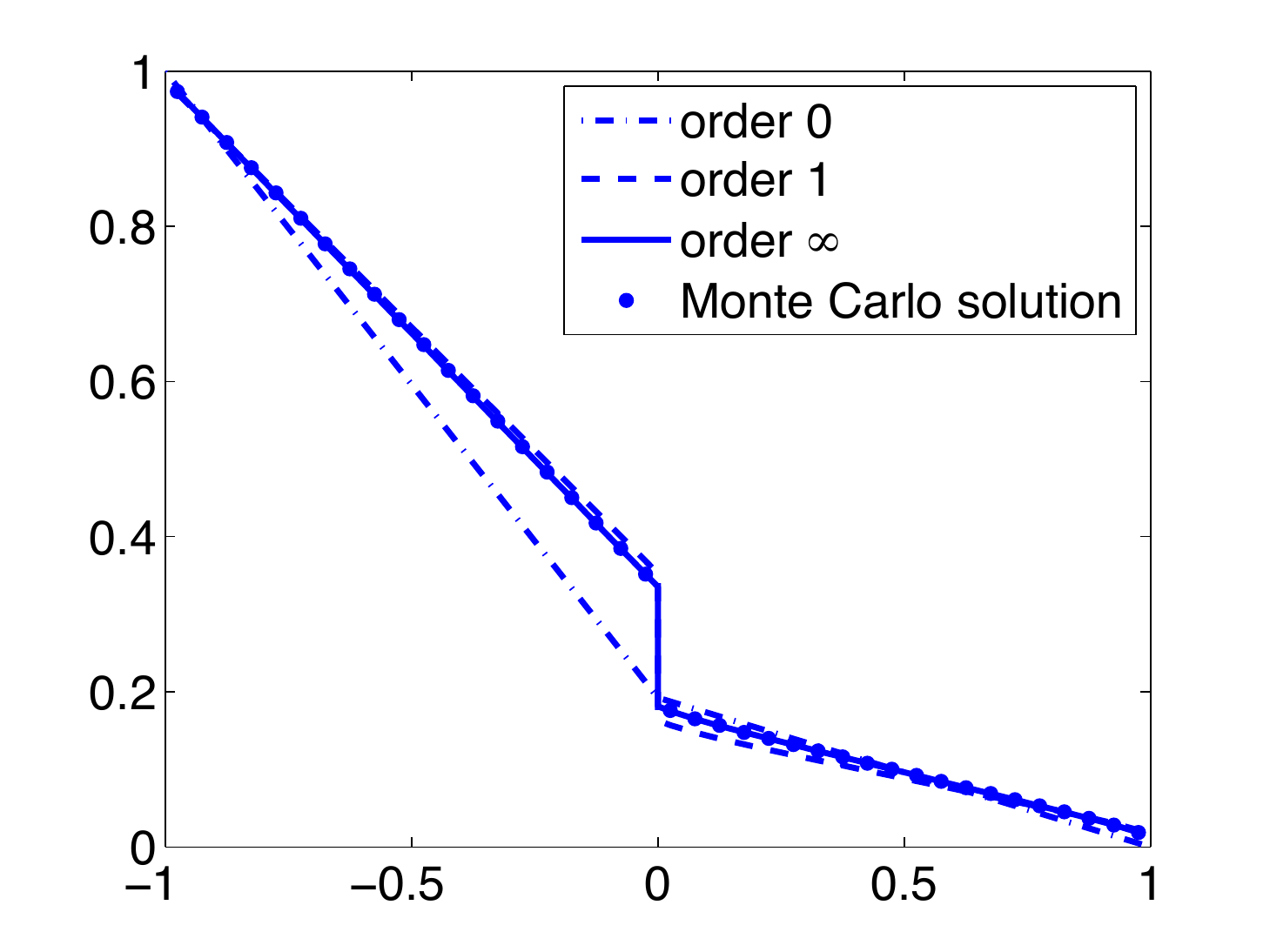}
	\caption{Temperature profile in a 1D system with an Al/Si interface}
		\begin{tikzpicture}[overlay]
        		\draw[](1,1.2) node[left]{\rotatebox{0}{ \large $x_1$ }};
        		\draw[](-3,3.8) node[left]{\rotatebox{90}{\small  $\displaystyle$ $T$ $(K)$}};
	        
	\end{tikzpicture}
	\label{fig:interface_asymptotic}
\end{figure}

Our numerical results are shown in Figure \ref{fig:interface_asymptotic}. The figure compares the temperature profile obtained with the deviational Monte Carlo method \cite{peraud12,Peraud2014Adjoint} to the order 0, order 1 and ``infinite'' order asymptotic solution. The order 1 solution provides significant improvement with respect to order 0. 
After adding the corresponding boundary layer functions we find that the infinite order solution agrees very well with the Monte Carlo result. Using this model, we obtain the actual conductance value $G=108$ MWm$^{-2}$K$^{-1}$, which is very close to the ``target'' value 110 MWm$^{-2}$K$^{-1}$ used as input to the {\it model} described in \cite{minnichthesis}. Perhaps more importantly, we note that the MC simulation also predicts a conductance value (obtained by extrapolating the bulk temperature profiles in order to calculate the temperature difference at the interface) of 108 MWm$^{-2}$K$^{-1}$, which is in perfect agreement with the (infinite order) asymptotic result. By comparison, the diffuse mismatch model predicts an interface conductance of $G = 343$ MWm$^{-2}$K$^{-1}$. This is consistent with the fact that the diffuse mismatch model results in an upper bound for the interface conductance \cite{zeng01}. 

We note that the ``infinite'' order solution may not be available in the general, higher-dimensional case. Related treatments of ``connection'' problems associated with different carriers have appeared in \cite{degond1998, aoki2008, aoki2007, takata2007}.

\section{Final remarks}
\label{final}
We have presented an asymptotic solution of the Boltzmann equation in the small Knudsen-number limit. The resulting solution provides governing equations and boundary conditions that determine the continuum temperature and heat flux fields in arbitrary three-dimensional geometries. Our results show that, for steady problems, the equation governing the {\it bulk} temperature field up to second order in the Knudsen number is the steady heat conduction equation. We also show that, up to first order in the Knudsen number, the equation governing the {\it bulk} temperature field in transient problems is the transient heat conduction equation. 

Although this result is expected (at least to first order in the Knudsen number) courtesy of traditional kinetic theory analysis \cite{Ziman1960,chen} (expanding the distribution function about the {\it local} equilibrium and giving no consideration to boundaries), the present work additionally derives the boundary conditions that complement this equation so that the resulting solutions of this system are {\it rigorously} consistent with solutions of the Boltzmann equation. In particular, the present work shows that the constitutive relation is only valid in the bulk, while a few mean free paths from the boundaries kinetic effects are always present. These effects not only modify the local constitutive relation (which is no longer of the Fourier-type), they also have a significant effect on the bulk solution by modifying the effective boundary condition subject to which the heat conduction equation is to be solved. These effective boundary conditions are derived for a variety of kinetic boundary conditions and shown to generally be of the jump type thus explaining the temperature jumps at the boundaries previously observed and remarked upon \cite{chen, lacroix05,yang2005}. We note here that the jump conditions are universal (non-adjustable), while the jump coefficients and kinetic boundary layers are universal for a given material and material-interface interaction model; in other words, they are independent of system dimensionality and once calculated they  can be used in any geometry of interest. Tabulated data for the various boundary layers derived not available in analytical form are available upon request.

These results provide no evidence or justification for modifying the material constitutive relation (thermal conductivity) as a means of extending the applicability of the traditional continuum description to the transition regime--the underlying physics is considerably more complex. According to the asymptotic theory presented here, in the regime $\langle \text{Kn}\rangle <1$ (strictly speaking $\langle \text{Kn}\rangle \ll 1$) solutions consistent with the Boltzmann equation are obtained using a thermal conductivity that is equal to the bulk value; the modified (typically reduced) transport rate associated with size effects {\it due to boundary presence} is captured by the additional resistance introduced by the jump boundary conditions as well as kinetic corrections that are to be linearly superposed to the final heat conduction result. On the other hand, by virtue of the expansion considered here, this work pertains to breakdown and extension of the classical Fourier description due to the inhomogeneity introduced by boundaries. As a result, it does not treat kinetic effects appearing in a spatially homogeneous material such as those arising from temporal variations that are fast compared to the relaxation time, or spatial variations that have characteristic lengthscales that are on the order of, or smaller than, the phonon mean free path, that are also of interest to the scientific community \cite{collins2013, hua2014b, Hu_2015}. Ultimately, a theory that captures both classes of kinetic effects under a unified framework needs to be developed; we hope that this work is a step towards that goal.

Our results are extensively validated using deviational Monte Carlo simulations of multidimensional problems.  
Studies in rarefied gas dynamics \cite{pof2006} show that second-order asymptotic formulations are reliable to engineering accuracy up to $\text{Kn}\approx 0.4$ and in some cases, depending on the problem simplicity, beyond. Our numerical validations support this finding. 

We note that the theory presented here assumes boundaries to be flat (no curvature). Curvature effects are expected to introduce 
additional terms in the effective boundary condition expressions and associated boundary layer corrections \cite{Sone2007}. This will be the subject of future work.

Due to its ability to capture the inhomogeneity in the distribution function associated with presence of boundaries, the present theory lends itself naturally to the description of the 
Kapitza resistance {\it and temperature jump} associated with the interface between two materials. We have shown that the asymptotic description produces results that are in excellent agreement with deviational Monte Carlo simulations. In other words, given transmission and reflection coefficients at the interface, the asymptotic theory may be used to {\it predict} the Kapitza resistance without any assumption on the form of the distribution in the interface vicinity.

\begin{appendix}
\section{Derivation of the governing equation for the order 1 and order 2 bulk temperature fields}
\label{laplace_demo}
In this section, we show that $T_{G1}$ and $T_{G2}$ are solution to the Laplace equation. We start with the case of $T_{G1}$. We apply the solvability condition \eqref{eq:solvability} to $\Phi_{G2}$ to obtain:
\begin{equation}
\int_{\omega,p,\boldsymbol{\Omega}}C_{\omega,p} V_{g} \boldsymbol{\Omega} \cdot \nabla_\mathbf{x} \left( T_{G2}-\frac{\text{Kn}}{\langle \text{Kn} \rangle}\boldsymbol{\Omega} \cdot \nabla_\mathbf{x} T_{G1} + \frac{\text{Kn}^2}{\langle \text{Kn} \rangle^2} \boldsymbol{\Omega} \cdot \nabla_\mathbf{x} \left( \boldsymbol{\Omega} \cdot \nabla_\mathbf{x} T_{0} \right)\right) d\omega d^2\boldsymbol{\Omega}=0.
\end{equation}
Integration over $d^2\boldsymbol{\Omega}$ removes terms containing odd powers of $\Omega_i$, yielding 
\begin{equation}
\int_{\omega,p,\boldsymbol{\Omega}}C_{\omega,p} V_{g} \frac{\text{Kn}}{\langle \text{Kn} \rangle} \sum_{i=1}^{3} \Omega_i^2 \frac{\partial^2 T_{G1}}{\partial x_i^2}  d\omega d^2\boldsymbol{\Omega}=0
\end{equation}
from which we conclude that 
\begin{equation}
\nabla_\mathbf{x}^2 T_{G1}=0
\end{equation}

To obtain the Laplace equation for $T_{G2}$, we apply \eqref{eq:solvability} to $\Phi_{G3}$. After carrying out the angular integration and cancelling terms containing odd powers of $\Omega_i$ we are left with
\begin{equation}
\int_{\omega,p,\boldsymbol{\Omega}}C_{\omega,p} V_{g}\left( \frac{\text{Kn}}{\langle \text{Kn} \rangle} \Omega_i^2 \frac{\partial^2 T_{G2}}{\partial x_i^2}  + \frac{\text{Kn}^2}{\langle \text{Kn} \rangle^2} \sum_{i,j,k,l} \Omega_i \Omega_j \Omega_k \Omega_l \frac{\partial^4  T_{0}}{\partial x_i \partial x_j \partial x_k \partial x_l} \right)d\omega d^2\boldsymbol{\Omega}=0.
\end{equation}
Thus, in order to show that the Laplace equation holds for $T_{G2}$, we need to show that 
\begin{equation}
\int_{\boldsymbol{\Omega}} \sum_{i,j,k,l} \Omega_i \Omega_j \Omega_k \Omega_l \frac{\partial^4  T_{0}}{\partial x_i \partial x_j \partial x_k \partial x_l}d^2\boldsymbol{\Omega}=0.
\end{equation}
Performing the angular integration, we obtain
\begin{equation}
\int_{\boldsymbol{\Omega}} \sum_{i,j,k,l} \Omega_i \Omega_j \Omega_k \Omega_l \frac{\partial^4  T_{0}}{\partial x_i \partial x_j \partial x_k \partial x_l}d^2\boldsymbol{\Omega}=
\frac{4 \pi}{5} \sum_{i,j}\frac{\partial^4 T_{0}}{\partial x_i^2 \partial x_j^2}=\frac{4 \pi}{5}\nabla_\mathbf{x}^2 \nabla_\mathbf{x}^2 T_{0}=0
\end{equation} 
as desired.

It appears that this procedure can be applied to all higher order terms ($T_{G3}$, $T_{G4}$, etc.). 

\section{Determination of jump coefficients and boundary layer functions in eqs \eqref{Order2Coefficients} and \eqref{Order2Functions}}
\label{Second order prescribed}
Coefficients $d_i$ and $g_{ij}$ (in \eqref{Order2Coefficients})and functions $\Psi_{K2,i}$ and $\Psi_{K2,ij}$ (in \eqref{Order2Functions}) are determined by boundary value problems of the same form as the ones discussed in section \ref{boundary_conditions} satisfying equation \eqref{eq:K1_boltz}.
The problems that determine the coefficients $\tilde{g}_{ij}$ include the source terms from that appear on the RHS of \eqref{second_order_BLequation}. In the interest of brevity, we only discuss the ones associated with the source term $-\partial \Phi_{K1}/\partial x_2$. The remaining two (associated with the term $-\partial \Phi_{K1}/\partial x_3$) may be deduced by analogy. 

For $i=2,3$ coefficient $\tilde{g}_{2i}$ is solution to:
\begin{equation} \label{BLO2_math_problem1}
\left \{
\begin{split}
&\Omega_1 \frac{\partial \tilde{\Psi}_{K2,2i}}{\partial \eta} = \frac{\langle \text{Kn} \rangle}{\text{Kn}} \left( \mathcal{L}(\tilde{\Psi}_{K2,2i})-\tilde{\Psi}_{K2,2i} \right) +\frac{\text{Kn}}{\langle \text{Kn} \rangle} \Omega_2 \Omega_i \exp \left( \frac{-\eta \langle \text{Kn} \rangle }{\Omega_1 \text{Kn} } \right)H(\Omega_1) \\
&\tilde{\Psi}_{K2,2i}(\boldsymbol{\Omega},\omega,p,\eta=0) +\tilde{g}_{2i}=0, \quad \text{for} \ \Omega_1>0\\
&\lim_{\eta \to\infty} \tilde{\Psi}_{K2,2i}(\boldsymbol{\Omega},\omega,p,\eta) =0
\end{split}
\right.
\end{equation}
where $H$ denotes the Heaviside function. Two results can be obtained immediately:
\begin{itemize}
\item[-] Coefficients $d_1$, $d_2$ and $d_3$ are solutions to the same problems as $c_1$, $c_2$ and $c_3$. Consequently, they are equal and their associated boundary layers are the same, provided $T_{0}$ is replaced by $T_{G1}$ in Eqs. \eqref{expression_K1i}, \eqref{BC1} and \eqref{BL1}. 
\item[-] Coefficients $g_{ij}$ and $\tilde{g}_{ij}$ for $i\neq j$ are zero.  For instance, it can be verified that 
\begin{equation}
\tilde{\Psi}_{K2,23}=
\left\{
\begin{split}
& \frac{\text{Kn}}{\langle \text{Kn} \rangle} \Omega_2 \Omega_3 \frac{\eta}{\Omega_1} \exp \left( \frac{-\eta \langle \text{Kn} \rangle}{\Omega_1 \text{Kn} }\right) \ \text{for} \ \Omega_1>0 \\
&0 \ \text{for} \ \Omega_1<0
\end{split}
\right.
\end{equation}
is a solution of  \eqref{BLO2_math_problem1}. Solutions for all $g_{ij}$ and $\tilde{g}_{ij}, i\neq j$ can be systematically obtained by solving the associated problem without the 
$\mathcal{L}(\tilde{\Psi}_{K2,ij})$ term and then verifying that $\mathcal{L}(\tilde{\Psi}_{K2,ij})=0$ 
\end{itemize}

We are left with five undetermined coefficients, namely $g_{11}$, $g_{22}$, $g_{33}$, $\tilde{g}_{22}$ and $\tilde{g}_{33}$. These can be determined using the numerical approach described in \cite{JPThesis} (suitably modified in order to accommodate the volumetric source terms which appear in the mathematical formulation). Instead of following this approach, here we prove that 
\begin{equation}
\sum_{i=1}^3 g_{ii} \left. \frac{\partial^2 T_{0}}{\partial x_i^2}\right|_{\eta=0}+\sum_{i=2}^{3} \tilde{g}_{ii} \left. \frac{\partial^2 T_{0}}{\partial x_i^2}\right|_{\eta=0}=0
\label{sum_of_2nd_order}
\end{equation}
and that, therefore, the temperature jump associated with the second order derivative is zero, while the boundary layer, although not zero, integrates into a zero temperature. 

The remaining five coefficients in relation \eqref{sum_of_2nd_order}, can be determined by finding the function  
$\tilde{\Psi}$ that satifies:
\begin{equation} \label{BLO2_eq_summary}
\left \{
\begin{split}
&\Omega_1 \frac{\partial \tilde{\Psi}}{\partial \eta} = \frac{\langle \text{Kn} \rangle}{\text{Kn}} \left( \mathcal{L}(\tilde{\Psi})-\tilde{\Psi} \right) +\sum_{i=2}^3\frac{\text{Kn}}{\langle \text{Kn} \rangle} \Omega_i^2 \left. \frac{\partial^2 T_{0}}{\partial x_i^2} \right|_{\eta=0}\exp \left( \frac{-\eta \langle \text{Kn} \rangle }{\Omega_1 \text{Kn} } \right)H(\Omega_1) \\
&\tilde{\Psi}(\boldsymbol{\Omega},\omega,p,\eta=0) =-\frac{\text{Kn}^2}{\langle \text{Kn} \rangle^2}\sum_{i=1}^{3} \Omega_i^2 \left. \frac{\partial^2 T_{0}}{\partial x_i^2} \right|_{\eta=0}, \quad \text{for} \ \Omega_1>0\\
&\lim_{\eta \to\infty}\tilde{\Psi}(\boldsymbol{\Omega},\omega,p,\eta) =0
\end{split}
\right.
\end{equation}
Let $\tilde{\Psi}\equiv \sum_{k=1}^5\tilde{\Psi}_k$, where $\tilde{\Psi}_k$ for $k=1,...,5$ correspond, respectively, to the five boundary layer functions that are the counterparts of the five temperature jump terms in relation \eqref{sum_of_2nd_order}.
We proceed with a strategy similar to the one used above, namely, 
solve for each $\tilde{\Psi}_k$ individually, ignoring the contribution of $\mathcal{L}(\tilde{\Psi}_k)$, and then evaluating $\mathcal{L}(\tilde{\Psi}_k)$. In the present case $\mathcal{L}(\tilde{\Psi}_k)\neq 0$ 
but $\sum_{k=1}^5\mathcal{L}(\tilde{\Psi}_k)=0$; more details can be found in \cite{JPThesis}. This proves that $\sum_{k=1}^5 \tilde{\Psi}_k$ is the solution of \eqref{BLO2_eq_summary} with the specified source terms and boundary conditions, and that the resulting boundary layer satisfies the boundary conditions without requiring a temperature jump correction, that is, relation \eqref{sum_of_2nd_order} is proved.

\end{appendix}

\bibliographystyle{asmems4}

\section*{Acknowledgment}
The authors would like to thank Professor T.R. Akylas for many helpful comments and discussions. N.G.H would also like to thank K. Aoki and S. Takata for many useful discussions. The preparation of this manuscript as well as the work on extension to time-dependent problems (section \ref{time-dep}) and the conductance of the interface between two materials (section \ref{kapitza}) was supported by the Solid-State Solar-Thermal Energy Conversion
Center (S3TEC), an Energy Frontier Research Center funded by the U.S. Department
of Energy, Office of Science, Basic Energy Sciences under Award\# DE-SC0001299 and DE-FG02-09ER46577. The remainder of the work was 
supported by the Singapore-MIT Alliance.

%

\small{

\bibliography{References}

}

%
%

\end{document}